\documentclass[12pt]{article}
\usepackage{amsmath,amsfonts,amssymb,amsthm}
\usepackage{nicefrac}
\usepackage{cite}
\usepackage{braket}
\usepackage{physics}

\usepackage{times}

\usepackage{xcolor}

\usepackage{geometry}
\usepackage{hyperref}

\newcommand{\poubelle}[1]{}

\def\CGHS{\widehat{\text{CGHS}}}
\newcommand{\x}{{\bf x}}

\usepackage[utf8x]{inputenc}
\usepackage[T1]{fontenc}
\usepackage{amsthm}
\usepackage{slashed}
\usepackage{mathtools}
\usepackage{float}
\usepackage{empheq}
\usepackage{bm}
\usepackage{framed}
\usepackage{comment}
\usepackage{xcolor}

\def\sss{\subsubsection}

\def\ie{\emph{i.e.} }

\def\nt{\notag}

\def\R{\mathbb{R}}

\def\cE{\mathcal{E}}
\def\cF{\mathcal{F}}

\def\cN{\mathcal{N}}
\def\cO{\mathcal{O}}
\def\cP {\mathcal{P}}

\def\p{\partial}

\def\/{\over}
\def\ov{\over}

\def\rn{\rangle}
\def\ln{\langle}

\def\e{\epsilon}
\def\ve{\varepsilon}
\def\vphi{\varphi}
\def\a{\alpha}
\def\b{\beta}
\def\d{\delta}
\def\k{\kappa}
\def\g {\gamma}
\def\la {\lambda}

\def\l{\ell}
\def\mn{{\mu\nu}}
\def\rs{{\rho\sigma}}

\def\n {\nabla}
\def\L{\Lambda}

\def\Om {\Omega}

\def\ra{\rightarrow}

\def\Tr{\mathrm{Tr}}

\def\Re	{\mathrm{Re}\,}
\def\r{\mathrm}

\def\_{\hspace{2cm}}
\def\-{\\\notag}
\def\={&=&}

\newcommand\be{\begin{equation}}
\newcommand\ee{\end{equation}}

\newcommand{\bea}{\begin{eqnarray}}
\newcommand{\eea}{\end{eqnarray}}

\newcommand{\bpm}{\begin{pmatrix}}
\newcommand{\epm}{\end{pmatrix}}

\newcommand{\bit}{\begin{itemize}}
\newcommand{\eit}{\end{itemize}}

\newcommand{\ben}{\begin{enumerate}}
\newcommand{\een}{\end{enumerate}}

\newcommand\bsp{\begin{split}}
\newcommand\esp{\end{split}}

\def\le{\left}
\def\ri{\right}

\def\ms{\medskip}

\def\l{\ell}

\def\qq{\qquad}

\def\Re{\r{Re}}

\newcommand{\subf}[2]{%
  {\small\begin{tabular}[t]{@{}c@{}}
      #1\\#2
    \end{tabular}}%
  }

\numberwithin{equation}{section}

 \title{Baby universes and near horizon dynamics from the CGHS model}

\author{}

\hypersetup{
    colorlinks=true,
    linkcolor=blue,
    filecolor=magenta,      
    urlcolor=cyan,
    citecolor = blue,
}

\begin{document}

\hfuzz=1.5pt





\def\mytitle{From black holes to baby universes \\[4mm] in CGHS gravity}

\pagestyle{myheadings} \markboth{\textsc{\small Godet, Marteau}}{%
  \textsc{\small }} \addtolength{\headsep}{4pt}

\begin{flushright}\small
\end{flushright}

\begin{centering}

  \vspace{1cm}

  \textbf{\LARGE{\textbf{\mytitle}}}



  \vspace{1.5cm}

  {\large Victor Godet$^{a}$ and Charles Marteau$^{b}$}

\vspace{1.5cm}

\begin{minipage}{1.0\textwidth}\small \it 
\begin{center}
${}^a$ International Centre for Theoretical Sciences (ICTS-TIFR),
Tata Institute of Fundamental Research, Shivakote, Hesaraghatta, Bangalore 560089, India\\
e-mail: victor.godet@icts.res.in\\
\vspace{0.3cm}
$^b$ Department of Physics and Astronomy,
University of British Columbia, Vancouver, BC V6T 0C2, Canada\\
e-mail: cmarteau@phas.ubc.ca
\end{center}
\end{minipage}

\end{centering}

\vspace{1cm}

\begin{center}
  \begin{minipage}{.9\textwidth}
    \textsc{Abstract}. 
     We study $\CGHS$ gravity, a variant of the matterless Callan-Giddings-Harvey-Strominger model. We show that it describes a universal sector of the near horizon perturbations of non-extremal black holes in higher dimensions. In many respects this theory can be viewed as a flat space analog of Jackiw-Teitelboim gravity. The result for the Euclidean path integral implies that $\CGHS$ is dual to a Gaussian ensemble that we describe in detail. The simplicity of this theory allows us to  compute exact quantities such as the quenched free energy and provides a useful playground to study baby universes, averages and factorization. In particular we derive a ``wormhole = diagonal'' identity.  We also give evidence for the existence  of a non-perturbative completion in terms of a matrix model. Finally, flat wormhole solutions in this model are discussed.

  \end{minipage}
\end{center}

\vfill

\thispagestyle{empty}
\newpage

\begingroup
\hypersetup{linkcolor=black}
\tableofcontents
\endgroup

 \newpage

\section{Introduction}

Quantum black holes offer a glimpse into the inner workings of quantum gravity. Recently, a theory of dilaton gravity in two dimensions, known as Jackiw-Teitelboim (JT) gravity \cite{Jackiw:1984je,Teitelboim:1983ux}, has been a useful tool in pushing forward our understanding of quantum black holes. This progress has happened on two different fronts as we have learned that there are actually two possible interpretations of JT gravity:
\bit
\item as an approximation of a higher-dimensional theory, which describes a universal sector of the near horizon dynamics of near extremal black holes \cite{Maldacena:2016upp}.
\item as a complete theory of quantum gravity in its own right, defined using the Euclidean path integral and equivalent to a random matrix ensemble \cite{Saad:2019lba}.
\eit
In this paper, we study a model of flat gravity in two dimensions, the $\CGHS$ model \cite{Afshar:2019axx}. We will see that it also has two possible interpretations, either as an approximate theory of the near horizon dynamics of \emph{non-extremal} black holes, or as a complete theory of \emph{flat gravity} in two dimensions. The first interpretation allows us to study the near horizon dynamics of generic black holes while the second gives us a simple model to study the gravitational path integral. In this introduction,  divided in two parts to reflect the two  interpretations mentioned above, we will review the context and briefly summarize the results of this work.

 \subsection{Black hole near horizon dynamics}

The enhancement of symmetries in the near horizon region of extremal black holes has been crucial in understanding their microscopic aspects \cite{Strominger:1996sh, Strominger:1997eq,Sen:2007qy}. However, because the AdS$_2$ geometry doesn't support finite energy excitations, the exact near horizon theory is a theory of ground states with no dynamics \cite{Maldacena:1998uz,Sen:2008vm}. More recently, it was understood that the deviation away from AdS$_2$ is governed by a two-dimensional dilaton gravity theory known as Jackiw-Teitelboim (JT) gravity and controlled by a universal pattern of symmetry breaking  \cite{Almheiri:2014cka,Jensen:2016pah,Engelsoy:2016xyb,Maldacena:2016upp}. This  has opened the way to study dynamical aspects of quantum black holes such as the information paradox \cite{Almheiri:2020cfm}.

We will propose here an analogous story for non-extremal black holes. Despite the fact that, in this case, the near horizon limit is not a decoupling limit, the near horizon region still has universal features such as the appearance of two-dimensional Rindler spacetime. We will see that the near horizon dynamics is also governed by a universal effective theory, $\CGHS$ gravity \cite{Afshar:2019axx}, which is also controlled by a symmetry breaking pattern.

In this introduction, we will first review how JT gravity arises in the near horizon region of near-extremal black holes and contrast it with our non-extremal story. For illustrative purpose, we will use the example of the BTZ black hole but the discussion applies more generally. 

\ms

Let us consider 3d Einstein gravity with a negative cosmological constant. We would like to study perturbations of a BTZ black hole. To do this, we write the following ansatz for a 3d metric 
\be\label{intro:BTZKKansatz}
g_3 = \hat{g}(x) + \hat\Phi(x)^2 (d\vphi + \hat{A}(x))^2~,
\ee
where hatted fields are two-dimensional and depend only on $x = (t,r)$. By dimensional reduction on the circle, we obtain a 2d effective theory \cite{Ghosh:2019rcj}
\be\label{introBTZaction}
I = -{1\/8G} \le[\int_M d^2 x \sqrt{\hat{g}}\,\hat\Phi\le( \hat{R}- {1\/4}\hat\Phi^2 \hat{F}_\mn \hat{F}^\mn + {2\/\l_3^2} \ri) + 2 \int_{\p M} ds\sqrt{\hat{h}}\,\hat\Phi \le(\hat{K}-{1\/\l_3}\ri)\ri]~.
\ee
This theory is fairly complicated. We will see that there are two different ways to make it simpler: the first applies to near-extremal black holes and leads to JT gravity while  the second applies to generic black holes and leads to $\CGHS$ gravity.

Starting with the near-extremal BTZ geometry, we can consider a near horizon limit given by 
\be
r\ra r_0 +\la r ,\qq t\ra {t\/2\la },\qq \vphi \ra \vphi + {t\/2\la} ~,
\ee
and taking $\la\to 0$. At the same time, we can deform the geometry using $r_\pm = r_0+\la\tau$. Here, $r_0$ is the position of the horizon at extremality and $\tau$ is a fixed constant. The limit $\la\to 0$ is a decoupling limit which leads to the near horizon geometry AdS$_2\times S^1$, where $\tau$ is ($2\pi$ times) the AdS$_2$ temperature. With the metric in the form \eqref{intro:BTZKKansatz}, the above procedure gives a particular expansion in $\la$ of the fields $\hat{g},\hat\Phi$ and $\hat{F}$, associated to the BTZ geometry. We can describe a more general near horizon perturbation by taking the same expansion in $\la$ but with arbitrary functions, so that at leading order
\be\label{introJTexpansion}
\hat{g} = g + O(\la),\qq \hat\Phi = \Phi_0+ \la \Phi+ O(\la^2)~.
\ee
The dynamics of these fields is described by the action \eqref{introBTZaction} evaluated on \eqref{introJTexpansion} which gives JT gravity
\be
I_\r{JT} = - S_0 \chi  - {\la \/8 G}\int d^2 x\sqrt{g} \,\Phi (R+2) + O(\la^2) + \text{boundary term}~,
\ee
where $S_0$ is the extremal entropy and the gauge field has been integrated out. This derivation was done in \cite{Ghosh:2019rcj} to which we refer for  more details. 

JT gravity was studied in \cite{Jensen:2016pah,Engelsoy:2016xyb,Maldacena:2016upp} where it was understood that the dynamics of the theory is controlled by symmetries, and admits an effective boundary description given by the Schwarzian action \cite{Kit.KITP.2,Maldacena:2016hyu}. The existence of a JT sector is expected to be universal for near-extremal black holes \cite{Almheiri:2014cka, Maldacena:2016upp,Almheiri:2016fws,Nayak:2018qej,Moitra:2019bub}, even though the near horizon dynamics can become more intricate in the presence of rotation \cite{Anninos:2017cnw,Castro:2018ffi,Castro:2019crn,Castro:2021csm}. This theory has led to insights into microscopic aspects of black holes, such as as the (non-)existence of a mass gap for (non-)supersymmetric near-extremal black holes \cite{Iliesiu:2020qvm,Heydeman:2020hhw}. 


\ms

In this work, we consider a different near horizon expansion.  We keep the black hole parameters general and take the near horizon limit
\be
r \ra r_+ +\la r~,
\ee
with no redefinition of $t$ or $\vphi$. The strict limit $\la\to0$ is singular as there is no decoupling limit for non-extremal black holes. At first order in $\la$, the near horizon geometry is 2d Rindler spacetime at the Hawking temperature, as is well-known.

As above, we consider more general perturbations by using the same expansion in $\la$ but with arbitrary functions, which gives
\be
\hat{g} = \la g + O(\la^2),\qq \hat\Phi = r_+(1+ \la \Phi)+ O(\la^2)~,\qq \hat{A} = {\la\/r_+} a + O(\la^2)~.
\ee
We see that the difference with \eqref{introJTexpansion} is the factor of $\la$ in the 2d metric, which reflects the absence of a decoupling limit. Still, we can study the near horizon dynamics using the action \eqref{introBTZaction} and we obtain the near horizon theory
\be
I = - S_0 \chi -{\la r_+\/8G}\int_M d^2x \sqrt{g}\le(\Phi R - {1\/4} f_\mn f^\mn +{2\/\l_3^2}\ri) + O(\la^2)+\text{boundary term}~. 
\ee
where $f= d a$ is the field strength associated to $a$ and $S_0$ is the entropy of the (non-extremal) unperturbed black hole. The term linear in $\la$ corresponds to the matterless CGHS model coupled to a $\mathrm{U}(1)$ Maxwell field. In Sec.~\ref{Near horizon dynamics of non-extremal black holes}, using a field redefinition, we show that this theory is equivalent to  $\CGHS$ gravity.

The  $\CGHS$ entropy, which is obtained from the appropriate boundary action, is given by the value of the dilaton at the horizon
\be
S_{\CGHS} ={\Phi|_{r=r_h}\/4 G_2}~.
\ee
where $G_2= G/(2\pi r_+) $ is the 2d Newton constant. This implies that $\CGHS$ correctly captures the change of black hole entropy.

Ultimately, we expect that $\CGHS$ gravity always appears   near the horizon of non-extremal black holes. Additional fields could appear when we consider more general black holes and theories  but the near horizon dynamics should always contain a $\CGHS$ sector, responsible for the linearized change in entropy. This makes it  an interesting theory to investigate. In particular, viewed  as a complete theory in its own right, one can study its gravitational path integral and its dual ensemble description, in analogy with the matrix model description of JT gravity.   The simplicity of $\CGHS$ makes it a useful playground to study various aspects of the Euclidean path integral in gravity.

\subsection{Euclidean path integral in gravity}

Several recent developments have improved our understanding of the gravitational path integral. One of the main question is the role of topology fluctuations and their effect on gravitational observables. In the late 80's wormholes and baby universes were considered and shown to lead to a probabilistic interpretation of the physics for the parent universe \cite{Lavrelashvili:1987jg, Hawking:1987mz, Giddings:1987cg, giddins1988, GIDDINGS1989481, coleman1988}. The effect of the baby universes on the parent universe is to produce undetermined coupling constants which belong to an ensemble whose probability measure depends on the type of baby universes one includes. Evolving the full state, \emph{i.e.} the state of the parent universe in tensor product with a generic baby universe state, leads to an effective loss of unitarity in the parent universe. Only a particular set of states maintain an effective unitarity, the so-called $\alpha$-states. The lack of knowledge due to the ensemble is reflected in the fact that we don't know which $\alpha$-state we live in. In this picture, determining completely the coupling constants of our parent universe corresponds to a collapse of the wave function into a particular $\alpha$-state.

One of the main subject where the gravitational path integral has been extensively studied is black hole thermodynamics. In conventional quantum field theory one computes the thermal partition function by path integrating the fields on a Euclidean cylinder whose circumference is the temperature. Things are different in gravity because the spacetime itself fluctuates. In particular, if one sticks to the QFT rule and imposes the existence of a thermal circle everywhere, one would miss the contribution of the black hole to the partition function. This is because the black hole actually corresponds to a configuration where the thermal circle becomes contractible in the bulk, at the location of the horizon. It seems that the only sensible thing to do in gravity is to impose the existence of a thermal circle at the boundary of spacetime. This prescription famously reproduces the Bekenstein-Hawking entropy \cite{Gibbons:1976ue}.


The gravitational path integral allows to define quantities by summing over all geometries with a fixed set of boundary conditions, consisting for example of multiple thermal circles. One interpretation of these quantities  is that they compute correlations between the partition functions of an ensemble of theories. The non-triviality of these correlations is directly related to the existence of geometries that connect multiple boundaries at once, \emph{i.e.} Euclidean wormholes. If there is no secret mechanism to exactly cancel these contributions, they appear to be in contradiction with holography \cite{Maldacena:2004rf}. For recent work on this factorization issue, we refer to the (non-exhaustive) list of references \cite{Betzios:2019rds, Blommaert:2019wfy, Marolf:2020xie, Pollack:2020gfa, VanRaamsdonk:2020tlr, Betzios:2020nry, Giddings:2020yes,Blommaert:2020seb,Anous:2020lka,Chen:2020tes,Eberhardt:2020bgq,Stanford:2020wkf,McNamara:2020uza,Cotler:2020lxj,Garcia-Garcia:2020ttf, Marolf:2021kjc,Eberhardt:2021jvj}.


This duality between gravity and an ensemble of holographic theories resonates with the earlier works on baby universes that we mentioned above. Actually the relation between the two approaches was made more precise in \cite{Marolf:2020xie} where the dual ensemble is revisited in terms of a baby universe Hilbert space. In this approach, states and operators are defined as boundary conditions in the gravitational path integral. The thermal partition function can be seen as one of these operators and the gravitational path integral with $n$ boundaries corresponds to an $n$-point function of this operator evaluated in the no-boundary state. In this construction, the ensemble naturally arises when the no-boundary state is decomposed into a basis of eigenstates, known as $\alpha$-states, for the partition function operator. One of the most surprising result in this approach is the fact that when one allows for geometries that connect more than two boundaries at once, the baby universe Hilbert space turns out to be much smaller than expected. 

The construction of \cite{Marolf:2020xie} is very general and we would like to understand how it is realized in known models of gravity. As computing the full gravitational path integral for theories in more than three dimensions seems out of reach, we focus on two-dimensional theories of gravity where the computation is manageable. The case of JT gravity was extensively studied and the dual ensemble and corresponding  $\alpha$-states have been understood \cite{Saad:2019lba,Blommaert:2019wfy,Blommaert:2020seb}.

The model studied in this paper is the $\CGHS$ model who is a variant of the well known CGHS model \cite{Callan:1992rs},  initially introduced as a toy model for black hole evaporation in two dimensions. Integrating over the dilaton gives the constraint $R=0$ so this is a model of gravity in \emph{flat spacetime}. The $\CGHS$ model is obtained after setting the matter fields to zero, performing a Weyl rescaling and ``integrating in'' two additional fields. This model allows for solutions of arbitrary temperatures whereas the temperature is fixed by the cosmological constant in CGHS. In \cite{Afshar:2019axx}, the authors show that one can rewrite the dynamics of $\CGHS$ in terms of a boundary action, which is a flat analog of the Schwarzian action. They also show that this boundary action can be obtained as some special scaling limit of the  complex SYK model, suggesting a holographic relation between $\CGHS$ and complex SYK similar to the one between JT gravity and SYK.

\subsection{Summary}

We will now summarize the results of this paper. In Sec.~\ref{The CGHS model}, we give an alternative definition of $\CGHS$ in terms of covariant boundary conditions. This gives another way to derive the boundary action. We then review the computation of the gravitational path integral obtained in \cite{Godet:2020xpk}. 

In Sec.~\ref{Near horizon dynamics of non-extremal black holes}, we show that $\CGHS$ captures the near horizon dynamics of non-extremal black holes, as was sketched in the first part of the introduction. We perform a dimensional reduction for both the BTZ and Reissner-Nordström black holes and show that $\CGHS$ emerges in the near horizon region. This universal $\CGHS$ sector captures the linearized change in black hole entropy. We comment on the fact that this universal sector could provide an effective field theory understanding of the first law of black hole thermodynamics.

In Sec.~\ref{Dual ensemble and baby universes}, we study $\CGHS$ as a complete theory on its own right. We give a thorough description of the dual ensemble and its "third-quantized" formulation in terms of baby universes. In particular, we obtain a bulk description of the factorization in an $\alpha$-state as a consequence of a ``wormhole = diagonal'' identity in the bulk, similar to the one that was derived in JT gravity \cite{Blommaert:2019wfy,Blommaert:2020seb}. We also address some issues associated with the free energy. In \cite{Engelhardt:2020qpv}, a prescription to compute the quenched free energy in gravity was proposed in terms of a replica trick involving replica wormholes. It was noted that the annealed free energies for the $\CGHS$ model and JT gravity are non-monotonous. This appears unphysical as it leads to a negative thermodynamical entropy. It was proposed that the physical free energy should be the quenched free energy, obtained by including contributions from replica wormholes, and that it should be monotonous. However, the analytic continuation in the replica trick is ambiguous which prevented a definite conclusion. In $\CGHS$, we can sidestep this issue by directly computing the quenched free energy. We find that it is still non-monotonous. This leads to the proposal that in general, the non-monotonicity is not resolved by going from annealed to quenched but is simply due to the fact that the ensemble contains members whose density of states is continuous.

Finally, we study wormhole solutions in $\CGHS$ plus matter in Sec.~\ref{sec:WH}. We find a Euclidean wormhole solution with interesting properties,  similar to the Euclidean wormhole in JT gravity described in \cite{Garcia-Garcia:2020ttf}. In particular, there is a phase transition at low temperature between the wormhole phase and a disconnected phase with two black holes. We also attempt to construct an eternal traversable wormhole but encounter various difficulties, which are similar to the difficulties in constructing wormholes with Poincaré symmetry in higher-dimensional AdS \cite{Freivogel:2019lej}.

\newpage

\section{The $\CGHS$ model}
\label{The CGHS model}

 The CGHS model without matter is a simple two-dimensional theory of flat space gravity. An inconvenient feature of this model is that the temperature of any solution is fixed by the cosmological constant. The $\CGHS$ model, first introduced in \cite{Afshar:2019axx}, has the same dilaton gravity sector as CGHS but with additional fields so that $\CGHS$ allows for solutions with different temperatures. It was shown in \cite{Afshar:2019axx} that the dynamics of $\CGHS$ reduces to a boundary action, which can be viewed as the flat space version of the Schwarzian action. In this section, we propose an alternative derivation of the boundary action of $\CGHS$. We also  review the computation of the Euclidean path integral \cite{Afshar:2019tvp,Godet:2020xpk}.

\subsection{Review of the model}

The CGHS model was first introduced to study the information paradox in two dimensions \cite{Callan:1992rs}. Without matter, it constitutes a simple theory of flat gravity in two dimensions, described by the Euclidean action
\begin{equation}
I_{\text{CGHS}}=-\frac{\kappa}{2}\int d^2 x \sqrt{\tilde{g}}\,e^{-2\tilde{\phi}}\left(\tilde{R}+4(\tilde{\nabla}\tilde{\phi})^2+2\Lambda\right),
\end{equation}
where $\k = (8\pi G_2)^{-1}$. This action can be simplified using a field redefinition involving a Weyl rescaling of the metric. Defining $\Phi=e^{-2\tilde{\phi}}$ and $g=e^{-2\tilde{\phi}}\tilde{g}$, we obtain the following action
\begin{equation}\label{CGHSaction}
I_{\text{CGHS}}=-\frac{\kappa}{2}\int d^2 x \sqrt{g}\left(\Phi R+2\Lambda\right)~.
\end{equation}
The corresponding equations of motion are 
\begin{equation}
R=0\quad \text{and}\quad \n_\mu \n_\nu \Phi -g_\mn \Box\Phi +\Lambda \,g_\mn  =0.
\label{CGHSeom}
\end{equation}
The first equation tells us that the geometry must be flat. In order to capture only physical degrees of freedom, we need to gauge-fix the metric. In AdS one can use the Fefferman--Graham gauge, but it does not exist in flat space. An alternative is the Bondi gauge which is adapted to null rays in Lorentzian signature. In Euclidean signature, the metric becomes complex and takes the form
\begin{equation}
ds^2=2V(\tau, r)d\tau^2+2id\tau dr~,
\end{equation}
where the metric is flat for $V(\tau,r)=T(\tau)+ P(\tau)r$.

The thermal solution is a disk  described by the metric
\begin{equation}
ds^2=\frac{4\pi r}{\beta}d\tau^2+2id\tau dr.
\label{DiskMetric}
\end{equation}
where $\tau\sim \tau+\beta$. The corresponding solution for the dilaton can be found by solving the second equation of \eqref{CGHSeom} which gives
\begin{equation}
\Phi=c_1+c_2\, e^{2\pi i\tau/\beta}+c_3\,r e^{-2\pi i\tau/\beta}+\frac{\beta \Lambda r}{2\pi}\,.
\label{CGHSsol}
\end{equation}
As a side remark, one can show that the vector $\xi^\mu=\epsilon^{\mu\nu}\partial_\nu\Phi$ is always a Killing vector when the second equation of \eqref{CGHSeom} is satisfied. The $c_2$ and $c_3$ solutions correspond to the two boosts of the $\mathrm{ISO}(2)$ symmetry of the disk geometry, to which we will come back later. 

Imposing a Dirichlet boundary condition 
\begin{equation}
\Phi \underset{r\to+\infty}{\sim} \phi_r r\, ,
\end{equation}
where $\phi_r$ is some fixed constant, we find a relation between the temperature, the renormalized value of the dilaton and the cosmological constant
\begin{equation}\label{relbetaLambda}
\beta=\frac{2\pi \phi_r}{\Lambda}.
\end{equation}
Therefore, the temperature is fixed in this theory. This is a known result \cite{Afshar:2019axx, Maldacena:2019cbz, Stanford:2020qhm} which shows that CGHS should be modified if we want to have solutions with different temperatures.

This can be done by promoting $\Lambda$ to a field and considering the following action
\be\label{CGHShat}
I_{\widehat{\mathrm{CGHS}}} =- {\k\ov 2} \int d^2 x\sqrt{g}\le( \Phi R + 2 \Lambda- 2 \Lambda \ve^\mn \p_\mu A_\nu\ri) ~,
\ee
which was called $\CGHS$ by the authors of \cite{Afshar:2019axx}. The corresponding equations of motion are 
\begin{align}
R & = 0, & \n_\mu \n_\nu \Phi -g_\mn \Box\Phi  + \Lambda\,g_\mn  &=0~,\\
 \Lambda & = \r{const} ,&  \ve^\mn \p_\mu A_\nu & = 1~.
\label{EOMFLAT}
\end{align}
This reproduces the dynamics of CGHS except that the constant $\Lambda$ can take any value. As a result, this model allows for solutions with different values of the temperature.

\subsection{Boundary action}\label{boundary action}

The $\CGHS$ action should be supplemented by a boundary term that will depend on the variational problem.  The goal is to find boundary conditions such that the whole gravity dynamics will be described by boundary modes. Such boundary conditions were proposed in \cite{Afshar:2019axx} and the corresponding boundary action was derived using the BF formulation of the $\CGHS$ theory. In the following we propose an alternative derivation of the boundary action using a more ``covariant'' method.

Let us consider a flat geometry characterized by the value of the functions $P$ and $T$ that we chose to be constant for simplicity.\footnote{All the surfaces we are going to consider have constant representatives.} The metric is
\begin{equation}
ds^2=2(T_0+ P_0\,\tilde{r})d\tilde{\tau}^2+2id\tilde{\tau}d\tilde{r}.
\end{equation}
We now solve the equations of motion for generic values $P_0$ and $T_0$ for these constants, we obtain
\begin{equation}
\begin{split}
\Phi &=c_1+c_2\, e^{iP_0\tilde{\tau}}+c_3\, e^{-iP_0\tilde{\tau}}\left(\tilde{r}+\frac{T_0}{P_0}\right)+\frac{ \Lambda }{P_0}\,\tilde{r}\,\\
A &= \tilde{r} d\tilde{\tau} - d\tilde{g},
\end{split}
\label{CGHSsolT0P0}
\end{equation}
where $\tilde{g}$ is any function. 

We now impose the following boundary conditions
\be\label{bdyCond}
\Phi|_\p = {\phi_r\/\e},\qq A|_\p = {d\tau\/\e},\qq ( \Phi-i A^\mu\partial_\mu \Phi)|_\partial = \phi_h~,
\ee
where $\phi_h$ is a fixed constant. The first condition determines the position of the boundary, in terms of the cutoff $\e$ that should be sent to zero at the end of the computation. The second condition fixes $A|_\p$, the pullback of $A$ on the boundary. The third condition is necessary if we want all the integration constants to be fixed in the solution \eqref{CGHSsolT0P0}. 

Since the geometry must be a patch of flat geometry, there is a coordinate system $(\tau,r)$ in which the boundary is at $r= 1/\epsilon$, so that we have
\begin{equation}
A = (r + \mu(\tau))\,d\tau + O(r^{-1}) ~,
\label{Constr}
\end{equation}
and the third boundary condition guarantees that there is no term proportional to $dr$ at this order. In addition to \eqref{bdyCond}, we will also impose 
\be\label{CGHSadditionalcond}
\mu(\tau) = 0~.
\ee
This condition is a condition on the solution space that can be viewed as a choice of the particular patch we are looking at.

 In the new coordinate system, the metric is also in Bondi gauge but with new values of $P$ and $T$. We can deduce their new values in the following way: the diffeomorphisms that preserve the Bondi gauge are
\begin{equation}
\tau \to f(\tau),\quad r\to \frac{1}{f'(\tau)}(r+h(\tau)),
\label{BMS2}
\end{equation}
where $f(\tau+\beta)=f(\tau)+\beta$ and $h(\tau+\beta)=h(\tau)$. They form what is called the $\mathrm{BMS}_2$ asymptotic symmetry group, which is the analog of the $\mathrm{Diff}(S^1)$ symmetry in AdS$_2$. From this, we infer that there exists such a diffeomorphism that relates the coordinates $(\tau,r)$ to the coordinates $(\tilde{\tau},\tilde{r})$. In the $(\tau,r)$ coordinates, the metric takes the form
\begin{equation}
ds^2=2(T(\tau)+P(\tau)\,r)d\tau^2+2id\tau dr,
\label{MetricImage}
\end{equation}
where
\begin{equation}
\begin{split}
P(\tau) &=P_0\, f'(\tau)-\frac{i f''(\tau)}{f'(\tau)},\\
T(\tau) &=T_0\, f'(\tau)^2+P_0\, f'(\tau)h(\tau)+ih'(\tau)-\frac{ih(\tau)f''(\tau)}{f'(\tau)}.
\end{split}
\end{equation}

The constraint on the gauge field \eqref{Constr} and the two boundary conditions \eqref{bdyCond} we have imposed on the dilaton are going to translate into equations of motion for $f$ and $h$.

Let's start with the constraint on the gauge field. In the new coordinate system we want $A$ to be proportional to $d\tau$ which imposes $\partial_{\tilde{r}}\tilde{g}=0$. The gauge field becomes
\begin{equation}
A=(r+h(\tau)-\partial_{\tau}g)d\tau.
\end{equation}
where we have defined $g$ through $\tilde{g}=g\circ f^{-1}$. So we see that to satisfy the condition \eqref{Constr} we need  $h=g'$. We also impose $g$ to be single-valued on the circle, which can be seen as an additional restriction on the solution space.  This implies that $h$ cannot have a zero mode on the circle. Using $h=g'$ in the coordinate transformations \eqref{BMS2} preserves the semi-direct structure $\mathrm{Diff}(S^1)\ltimes C^\infty(S^1)$ but modifies the group law. As a result, the corresponding algebra becomes the warped Witt algebra. In addition, one can show that $P$ and $T$ transform in the coadjoint representation of the group \cite{Afshar:2019tvp,Afshar:2019axx}.

The boundary conditions on the dilaton lead to two equations. To obtain them we write our solution \eqref{CGHSsolT0P0} in the $(\tau,r)$ coordinates with the replacement $h=g'$. This gives
\begin{equation}
\begin{split}
\phi_r& =\frac{\Lambda}{P_0 f'}+c_3\frac{e^{-iP_0 f}}{f'}~,\\
\phi_h &=c_1+c_2 \,e^{iP_0f}+c_3\frac{T_0\,e^{-iP_0f}}{P_0}+\left(c_3\,\,e^{-iP_0f}+\frac{\Lambda}{P_0}\right)\frac{g'}{f'}~.
\end{split}
\end{equation}
These two constraints are the integrated versions of two differential equations that need to be satisfied by $f$ and $g$:
\begin{equation}
\begin{split}
P_0 f'-\frac{i f''}{f'}-{\Lambda\/\phi_r} &=0~,\\
2i T_0 f''+iP_0 g''+\frac{f''g''}{(f')^2}-\frac{g'''}{f'} &=0~.
\end{split}
\label{EOMbm}
\end{equation}
These two equations are of central importance as they control the dynamics of our two boundary modes $f$ and $g$. This is similar to the way the boundary conditions  constrains the boundary mode in JT gravity \cite{Maldacena:2016upp}. One thing to notice is that these two equations are independent of $\phi_h$, the constant that appears in the third boundary condition. This constant is actually equal to the value of the dilaton at the location of the horizon. Indeed,  one can check that the solution on the disk is 
\begin{equation}
\Phi= \phi_r r +\phi_h,
\end{equation}
and the horizon of  \eqref{DiskMetric} is located at $r=0$. This constant therefore sets the entropy of the thermal solution through the known relation $S=2\pi \kappa\,\Phi|_{r=r_h}=2\pi \kappa \phi_h$, also  derived in the next section.
We will see in Sec.~\ref{Near horizon dynamics of non-extremal black holes}, that this entropy can be seen as the entropy of the near horizon perturbation of a higher-dimensional black hole.

The dynamics described by \eqref{EOMbm} can also be recovered by computing the effective action for the boundary modes $f$ and $g$. In order to have a well defined variational problem, we need to add a boundary term to the action \eqref{CGHShat} which we take to be
\begin{equation}
I_\partial= -\kappa \int_{\partial} \sqrt{h} \left(\Phi K-\frac{1}{2}n^\mu\partial_\mu \Phi\right).
\end{equation}  
The first term makes the variational problem well-defined for Dirichlet boundary conditions on the metric and dilaton. The second term is a counterterm that does not affect the variational problem. Indeed from the third boundary conditions \eqref{bdyCond} we deduce that $(A^\mu \delta \partial_\mu \Phi)_\partial =0 $, and since $A^\mu \propto n^\mu$ on the boundary, we also have $(n^\mu \delta \partial_\mu \Phi)_\partial =0 $. Evaluating the full action on our solution space, the bulk term vanishes and the boundary term is finite, giving
\begin{equation}
I_\partial=\kappa \int_0^\beta d\tau\left(\phi_r\, T(\tau)-\phi_h \,P(\tau)\right).
\label{onshellaction}
\end{equation}
When writing $P$ and $T$ in terms of the two boundary modes (with $h=g'$) and the constants $P_0$ and $T_0$, we obtain an action for $f$ and $g$:
\begin{equation}
I_\partial[f,g]=\gamma\int_0^\beta d\tau \left(T_0 f'^2+P_0\, f'g'+i g''-\frac{ig'f''}{f'}\right)-\kappa \phi_h \beta P_0~,
\label{baction}
\end{equation}
where we have defined $\gamma\equiv\kappa\phi_r$. This is the action that was derived in  \cite{Afshar:2019axx} by a different method. Its equations of motion reproduce exactly the two equations \eqref{EOMbm}. When evaluated on the thermal solution with $P_0=2\pi /\beta$, this effective action has an $\mathrm{ISO}(2)\times \mathbb{R}$ global symmetry, otherwise  the symmetry is $ \mathbb{R}\times  \mathbb{R}$. The corresponding field transformations are 
\begin{equation}
\delta f= \lambda_1+\lambda_2\, e^{-iP_0 f(\tau)},\quad \delta g =\lambda_4+\lambda_3\,e^{iP_0 f(\tau)}-\lambda_2 \,\frac{T_0}{P_0}\,e^{-iP_0 f(\tau)}.
\label{Transfo}
\end{equation}
We can see that the generators $\lambda_2$ and $\lambda_3$ are well defined only when $P_0=2\pi  /\beta$.  The field transformations \eqref{Transfo} correspond to asymptotic diffeomorphisms in the bulk so we can associate surface charges to them. One can show that they match exactly the Noether charges of the boundary action \cite{Godet:2020xpk}.

This boundary action is an effective action for the Goldstone modes under the symmetry breaking from the warped Witt group $\mathrm{Diff}(S^1)\ltimes C^\infty(S^1)$  to the appropriate global symmetry, $\r{ISO}(2)\times \R$ for the disk and $\R\times \R$ for the cylinder. This leads to a pattern of spontaneous and explicit symmetry breaking similar to the Schwarzian story \cite{Maldacena:2016upp}. Another similarity is that this action can be interpreted as a coadjoint action of the warped Witt group \cite{Afshar:2019tvp} which implies that the path integral is one-loop exact \cite{Stanford:2017thb,Afshar:2019tvp}. It was also shown in \cite{Afshar:2019axx} that it arises from a particular scaling limit of the low-energy effective action of the complex SYK model, establishing a holographic relation between complex SYK and $\CGHS$.


\subsection{Euclidean path integral}
\label{Euclidean path integral}

Having found the boundary action, we can now use it to compute the Euclidean path integral. This computation was done in \cite{Godet:2020xpk} but will be briefly review it here for completeness. Our boundary conditions \eqref{bdyCond} select geometries that have thermal circles at infinity in the sense that the boundaries need to be at infinite distance from any point in the bulk. This excludes geometries like a flat disk with holes in it which would correspond to a configuration with multiple boundaries.\footnote{These geometries could appear in the finite cutoff version of the theory, see Fig.~\ref{FigFC} for illustration.} The integration over the dilaton constrains the metric to be flat. There are only two flat surfaces with asymptotic boundaries that are connected and regular: the disk and the cylinder\footnote{Indeed, there are only three classes of Riemann surfaces which can have a flat metric that is complete: the infinite disk, the cylinder and the torus. Allowing for incomplete metrics, which means that some boundaries will be at finite distance, more general surfaces are possible, such as the disk with holes depicted in Fig.~\ref{FigFC}. However these geometries should not be included because $\widehat{\text{CGHS}}$ is only defined for asymptotic boundaries.}. This will greatly simplify the computation. 

We consider the full gravitational action
\begin{equation}
I=I_{\mathrm{top}}+I_{\CGHS}+I_{\partial}\,,
\end{equation}
where we have added the topological term
\begin{equation}
I_{\mathrm{top}}=-\frac{S_0}{2\pi}\left[\frac{1}{2}\int_{\mathcal{M}}\sqrt{g} R+\int_{\partial \mathcal{M}}K\right],
\end{equation}
which controls the topological expansion. Consider the gravitational path integral on geometries with $n$ boundaries, formally written as a sum
\begin{equation}
Z_{\mathrm{grav}}(\beta_1,\ldots,\beta_n)=\sum_{g=0}^\infty \frac{Z_{g,n}(\beta_1,\ldots,\beta_n)}{(e^{S_0})^{2g+n-2}},
\end{equation}
since we have $I_{\mathrm{top}}=-S_0\chi=-S_0(2-2g-n)$. Now for each $Z_{g,n}$ the topology of the surface is fixed. Thanks to the constraint imposed by the integration over the dilaton, the path integral on fixed topology reduces to an integration over bulk moduli and boundary modes. We will have only two connected contributions to this sum: the disk and the cylinder. The first one gives us to the density of states while the second one tells us about correlations in the spectrum.

\paragraph{Disk.}

The disk contribution is obtained by computing the path integral of the boundary action \eqref{baction} on the disk geometry, which corresponds to the values:
\begin{equation}
P_0=\frac{2\pi }{\beta},\quad T_0=0.
\end{equation}
The corresponding Euler characteristic is $\chi=1$. The disk partition  function is one-loop exact and was computed in \cite{Afshar:2019tvp, Godet:2020xpk} leading to
\be
Z^{\r{disk}}(\b) = {2\pi\g^2\ov \b^2}e^{S_0+2\pi\kappa \phi_h} ~.
\label{ZDisk}
\ee
The term in the exponential corresponds to the classical part while the prefactor comes from the one-loop contribution. From the classical part we can read off the classical entropy 
\begin{equation}
S_{\mathrm{cl.}}=-(1-\beta \partial_\beta)I_{\mathrm{cl.}}=S_0+2\pi \kappa\phi_h.
\end{equation}
This justifies that the entropy of $\CGHS$ is given by the value of the dilaton at the horizon
\be
S_{\CGHS} = {\Phi|_{r=r_h}\/4 G_2} = 2\pi \k \phi_h~.
\ee
Having in mind the higher-dimensional picture, $S_0$ can be seen as the entropy of the unperturbed black hole while the term $2\pi \kappa \phi_h$ is the entropy added by the perturbation. In Sec.~\ref{Near horizon dynamics of non-extremal black holes}, $S_0$ will be the entropy of an unperturbed non-extremal black hole and will depend on $\beta$. Here and in Sec.~\ref{Dual ensemble and baby universes} we study the $\CGHS$ model as a 2d theory of gravity in its own right and take $S_0$ to be a constant for simplicity.

 An inverse Laplace transform allows us to extract a linear density of states 
\be\label{densityCGHS}
\rho^{\mathrm{disk}}(E) =2\pi \g^2 e^{S_0+2\pi\kappa \phi_h}  E~.
\ee
In the following, $\rho^{\mathrm{disk}}(E) $ will actually be interpreted as an average density of states $\langle \rho(E)\rangle$ in an ensemble that we will define. To probe the structure of this ensemble we also need to compute the correlation functions of $\rho(E)$, which will be contained in the cylinder contribution.

\paragraph{Cylinder.}

\begin{figure}
  \centering
  \includegraphics[width=8cm]{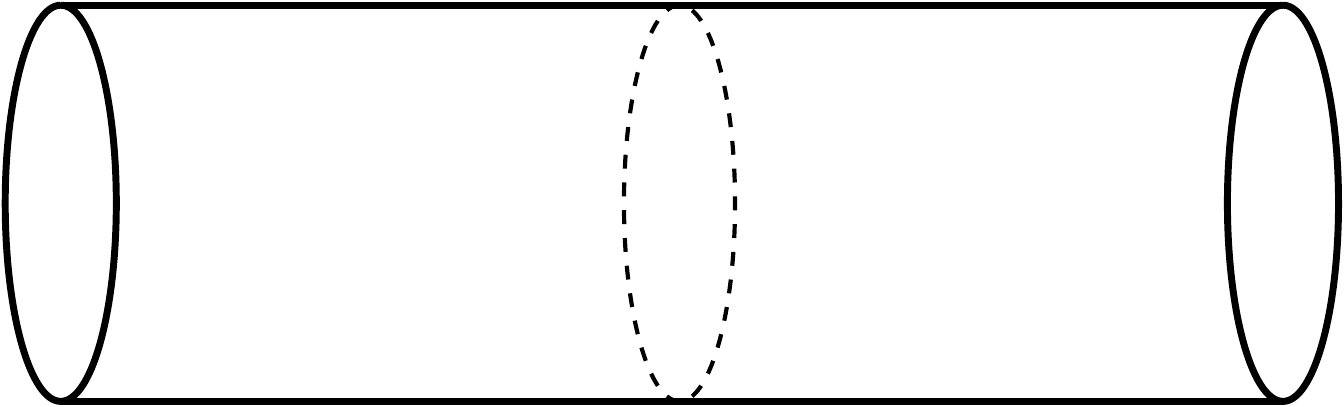}
  \put(-245,30){{$\b_1$}}
  \put(-115,75){{$b$}}
      \put(4,30){{$\b_2$}}
  \caption{The cylinder geometry. We specify two boundary conditions $\b_1$ and $\b_2$ at each end. The only modulus is the circumference $b$.}\label{Fig:Cylinder}
\end{figure}

The cylinder is the flat space analog of the double trumpet geometry in JT gravity. It has vanishing Euler characteristic and is described by the metric 
\be
ds^2 = dt^2 + d x^2~,\qq t\sim t +b~,
\ee
where $b>0$ is the circumference of the cylinder as depicted in Fig.\ \ref{Fig:Cylinder}. We first compute the contribution of half-cylinder which is described by the same metric, focusing only on the asymptotic boundary at $x=+\infty$. We can bring this metric to Bondi gauge with the following change of coordinates
\be
t= {b\ov \beta}\tau +{\beta\ov b} i r,\qq x = {\b\ov b}r ~.
\ee 
The metric becomes
\be
ds^2 = {b^2\ov \b^2} d\tau^2 + 2 i d\tau dr ~,
\ee 
which corresponds to the Bondi gauge with
\be
P_0 = 0,\qq T_0=  {b^2\ov 2 \b^2} ~.
\ee
The computation of the path integral for the half-cylinder is done in \cite{Godet:2020xpk} and gives
\be
Z^\text{half-cyl}(\b,b) = {2\pi \g\ov\b} \exp\le(-{\g b^2\ov 2\b} \ri) ~.
\ee
The result depends both on the temperature and the circumference of the half-cylinder. Again the path integral is one-loop exact: the term in the exponential corresponds to the classical action while the prefactor is the one-loop contribution.

\ms

To obtain the cylinder contribution, we glue two half-cylinders of equal circumference together as depicted in Fig.~\ref{Fig:Cylinder}. This leads to
\be
Z^\text{cyl}(\b_1,\b_2) =  \int_0^{+\infty} b db \, Z^\text{half-cyl}(\b_1, b) Z^\text{half-cyl}(\b_2,b)~,
\ee
where the factor of $b$ in the measure corresponds to the freedom to twist one of the half-cylinder relative to the other when gluing. Finally, we obtain
\be\label{Zcylinder}
Z^\text{cyl}(\b_1,\b_2)  = {4\pi^2\g\ov \b_1+\b_2}~.
\ee
\paragraph{Interpretation.} The fact that the cylinder contribution does not vanish means that the full gravitational path integral with two asymptotic boundaries does not factorize. This is in contradiction with the existence of a conventional holographic dual since in the dual picture, the path integral on two disconnected circles factorizes. A possible interpretation of this is that the gravitational path integral actually computes an ensemble average:
\begin{equation}
Z_{\mathrm{grav}}(\beta_1,\ldots \beta_n)=\langle Z(\beta_1)\ldots Z(\beta_n)\rangle~,
\end{equation}
and the fact that the cylinder contribution does not vanish corresponds to the non-factorization of the two-point function
\be
\ln Z(\b_1)Z(\b_2) \rn \neq \ln Z(\b_1) \rn\ln Z(\b_2)\rn~.
\ee
The connected $n$-point function correspond to connected flat surfaces with $n$ asymptotic boundaries. As such surfaces don't exist for  $n\geq 3$, we have
\be
\ln Z(\b_1)Z(\b_2)\dots Z(\b_n)\rn_c = 0 \qq n\geq 3~.
\ee
The vanishing of all the connected correlators for $n\geq 3$ implies that the third quantized theory should be Gaussian. This will be useful to construct the dual ensemble in Sec.~\ref{Dual ensemble and baby universes}.

\section{ Near horizon dynamics of non-extremal black holes}\label{Near horizon dynamics of non-extremal black holes}

We have seen that the $\CGHS$ model can be considered in many respects as a flat space analog of JT gravity. The near horizon dynamics of near-extremal black holes contains a universal sector described by JT gravity. In this section, we will see that  $\CGHS$ captures the near horizon dynamics of \emph{non-extremal} black holes. We will argue that there is a universal $\CGHS$ sector in the near horizon region which accounts for the linearized change in entropy.

\subsection{BTZ black hole}

A simple setting to study black hole perturbations is the BTZ solution of 3d gravity. We will show in detail how $\CGHS$ emerges as a two-dimensional effective theory describing the near horizon dynamics. The Reissner-Nordström black hole is studied in the next section.

\subsubsection{Near horizon expansion}

The BTZ geometry is described by the line element
\be
ds^2 = - {(r^2-r_+^2)(r^2-r_-^2)\/r^2} dt^2 + \l_3^2 {r^2 dr^2\/(r^2-r_+^2)(r^2-r_-^2)} + r^2 \le(d\vphi-{r_-r_+\/r^2}dt\ri)^2~,
\ee
where $r_\pm$ are the outer/inner horizons and $\l_3$ is the AdS$_3$ length. The mass and angular momentum  of the black hole are
\be
M = {r_+^2 + r_-^2\ov 8 G \l^2_3},\qq J = {r_+r_-\ov 4 G \l_3}~,
\ee
and the entropy, temperature and angular velocity at the horizon are 
\be
S = {\pi r_+\ov 2 G},\qq T = {1\/\b}= {r_+^2 - r_-^2 \ov 2\pi \l^2_3 r_+}~,\qq \Om_H = {r_-\ov r_+}~.
\ee
For our purposes, it will be convenient to use infalling coordinates obtained by replacing 
\be
t \ra {u\/\l_3} + r_\ast ,\qq \phi \ra\phi +r_\ast~, \qq d r_\ast \equiv {\l_3 r^2\/(r^2 -r_-^2)(r^2 - r_+^2)}dr~,
\ee
which leads to the metric
\be
ds^2 = -{(r^2- r_-^2)(r^2-r_+^2)\/\l_3^2 r^2 }du^2 - 2 du dr + r^2\le( d\vphi - {r_- r_+\/\l_3 r^2} du\ri)^2~.
\ee
This coordinate system is useful since the 2d metric in the near horizon region will be automatically in Bondi gauge. We now consider a near horizon limit
\be\label{BTZnearhorizon}
r \ra r_+ + \la r~,\qq (\la\to 0 )~.
\ee
with no redefinition of $t$ and $\vphi$. At the same time, we perform a deformation of the BTZ geometry:
\be\label{deformationBTZ}
r_+ \ra r_+ +\la\d r_+ + O(\la^2),\qq r_-\ra r_- -\la \d r_-+ O(\la^2)~,
\ee
controlled by the same small parameter $\la$. 

This procedure where we deform the geometry and go to the near horizon region is familiar from the study of near-extremal black holes. Here, we apply it to non-extremal black holes. We obtain the metric
\be\label{firstorderBTZ}
ds^2 = \la g_{\mn} dx^\mu x^\nu + (1+\la\Phi)^2 \le(r_+ d\vphi +d\a + \la a\ri)^2 + O(\la^2)~,
\ee
where $\a=-{r_-\/\l_3 } u $ and 
\bea\-
g_{\mn} dx^\mu x^\nu \= 2 (P_0 r + T_0 )du^2 -2 du dr~,\\\label{BTZexpansionValues}
\Phi  \= {r\/r_+}~,\-
a \= {2 r_-\/\l_3 r_+} r d u + {r_-\/\l_3 }\le({\d r_-\/r_-} - {\d r_+\/r_+} \ri) du ~,
\eea
with the values
\be
P_0 = -{2\pi \/\b},\qq  T_0  = {2\pi \d r_+\/\b}~.
\ee
This metric satisfies $R_{ab}+\frac{2}{\ell^2_3}g_{ab}=O(\lambda)$. As expected, the 2d geometry corresponds to a Rindler patch whose temperature matches the temperature of the black hole. Looking at the 2d metric, we see that the horizon is located at $r=-T_0/P_0=\delta r_+$.

An important difference with the near-extremal case is that there is no decoupling limit here. In particular, the metric becomes degenerate in the strict $\la\to 0$ limit because of the $\lambda$ factor multiplying the 2d metric. As a result, the leading order of the metric \eqref{firstorderBTZ} is \emph{not} an exact solution of Einstein's equation. We can obtain a higher order solution by including higher order terms in the metric. More precisely, the near horizon metric at order $\lambda^n$ will satisfy $R_{ab}+\frac{2}{\ell^2_3}g_{ab}=O(\lambda^{n-1})$. This is in contrast with the near-extremal case where at leading order, the near horizon metric is the AdS$_2\times S^1$ geometry which is an exact solution of Einstein's equation. Even though there is no decoupling limit here, we will see that the near horizon dynamics can be consistently studied.

\sss{Dimensional reduction}

The BTZ black hole is a solution of pure 3d gravity with a negative cosmological constant, described by the action
\be
I = -{1\/16\pi G}\le[\int_M d^2x d\vphi \sqrt{g_3} \le(R_3 +{2\/\l_3^2} \ri)  + 2\int_{\p M} d y d\vphi\sqrt{h} \le(K_3-{1\/\l_3}\ri) \ri]~,
\ee
where $x=(t,r)$ are 2d coordinates. Following  \cite{Ghosh:2019rcj}, we consider the Kaluza-Klein reduction of this theory using
\be\label{BTZKKansatz}
g_3(x,\vphi) = \hat{g}(x) + \hat\Phi(x)^2 (d\vphi + \hat{A}(x))^2~,
\ee
where we take the 2d metric $\hat{g}$, gauge field $\hat{A}$ and dilaton $\hat\Phi$ to only depend on $x$. We focus on the low energy dynamics near the horizon and ignore the massive Kaluza-Klein modes arising from the circle reduction. After integrating over the circle, the action becomes  \cite{Ghosh:2019rcj}
\be\label{BTZreducedaction}
I = -{1\/8G} \le[\int_M d^2\x \sqrt{\hat{g}}\,\hat\Phi\le( \hat{R}- {1\/4}\hat\Phi^2 \hat{F}_\mn \hat{F}^\mn + {2\/\l_3^2} \ri) + 2 \int_{\p M} ds\sqrt{\hat{h}}\,\hat\Phi \le(\hat{K}-{1\/\l_3}\ri)\ri]~,
\ee
where $\hat{F} = d\hat{A}$ is the field strength associated with the gauge field $\hat{A}$.

The resulting 2d theory is rather complicated. It was shown in \cite{Ghosh:2019rcj} that it greatly simplifies in the near-extremal limit, using a near horizon expansion of the fields, where it actually reduces to JT gravity. In this work, we keep the parameters of the black hole generic and we use a different near horizon expansion of the fields. 

The near horizon expansion of the deformed BTZ geometry given by \eqref{firstorderBTZ} with \eqref{BTZexpansionValues} can be viewed as a particular near horizon perturbation of the black hole. To capture more general perturbations, we will use the same expansion in $\la$ but with arbitrary functions. This leads to the following near horizon ansatz for the fields
\bea\nt
\hat{g}  \= \la \le(g+\la \g\ri) + O(\la^3)~,\\\label{expBTZ}
\hat{A} \={1\/r_+}\le( d\a   + \la a +\la^2 b \ri)+ O(\la^3)~, \-
\hat\Phi\= r_+(1+\la\Phi+\la^2\chi) + O(\la^3)~,
\eea
where the coefficients are arbitrary functions of the 2d coordinates. At first order in $\la$, this is the ansatz \eqref{firstorderBTZ} used in the previous section. We have also included fields of second order here. Again, the absence of decoupling limit is reflected in the degenerate nature of the $\la\to 0$ limit.  

It will be convenient to use a gauge in which the metric $\g$ is traceless: $\g_\mn g^\mn=0$. This can be achieved by putting $\la^{-1} \hat{g}$ in Bondi gauge. The near horizon expansion leads to  
\begin{align}
\sqrt{\hat{g}} & = \la \sqrt{g} + O(\la^2), & & 
\\
\hat{R} & =  \la^{-1} R + \le(\n_\mu \n_\nu- g_\mn \Box\ri)\g^\mn+ O(\la)~, & & \\\label{BTZexpF}
\hat{F}_\mn & = {\la\/r_+ }f_\mn+ O(\la^2) ~.
\end{align}
The expansion of the 2d reduced action \eqref{BTZreducedaction}  gives an expansion of the form
\be
I_\r{2d} = I_\r{2d}^{(0)}+\la I_\r{2d}^{(1)}+\la^2 I_\r{2d}^{(2)}+ O(\la^3)~.
\ee
and we find
\bea
I_\r{2d}^{(0)} \= -{r_+\/8G}\int_M d^2x \sqrt{g}\, R+  \text{boundary term}\\\label{2dtheoryBTZ}
I_\r{2d}^{(1)} \=  -{r_+\/8G}\int_M d^2x \sqrt{g}\le(\Phi R - {1\/4} f_\mn f^\mn +{2\/\l_3^2}\ri)+  \text{boundary term}~.
\eea
This action describes the dynamics close to the horizon. To properly define the variational problem, we need to introduce an imaginary boundary close to the black hole horizon, which plays the role of the asymptotic boundary of the near horizon theory.  This is the same procedure as used for example to derive JT gravity from the near-extremal BTZ black hole in \cite{Ghosh:2019rcj}. With the appropriate boundary term, the zeroth order term gives a topological contribution of the form
\be
I_\r{2d}^{(0)}=-\frac{S_0}{2\pi}\left[\frac{1}{2}\int_{\mathcal{M}}\sqrt{g} R+\int_{\partial \mathcal{M}}K\right]=-S_0 \chi,
\ee
where $S_0 = {\pi r_+\/2 G}$ is the entropy of the unperturbed black hole.\footnote{We use the notation $S_0$ here even though the black hole is not extremal.}

We would like now to study the dynamics described by the first order action $I_\r{2d}^{(1)}$. The equations of motion coming from the first order piece are
\bea
0 \= R  ~,\-
0\= \n_\mu f^\mn~,\-
0 \= \n_\mu\n_\nu\Phi - g_\mn\Box\Phi + {1\/\l_3^2}g_\mn + {1\/2}  f_{\mu\rho} f_\nu^{~\rho} -{1\/8}g_\mn f_\rs f^\rs  ~.
\eea
The first equation is a constraint that forces the 2d geometry to be flat. The equation for the gauge field implies that
\be\label{onshellFtilde}
f_\mn = 2 X \,\ve_\mn~,
\ee
where $X$ is a constant which can be interpreted as a two-dimensional electric field. The dilaton equation of motion then becomes
\be
\n_\mu\n_\nu\Phi - g_\mn\Box\Phi +g_\mn \L   =  0~,\qq \L \equiv X^2+{1\/\l_3^2} ~.
\ee
This shows that the near horizon dynamics is governed by the equations of motion of CGHS gravity \eqref{CGHSeom} with the only difference that the cosmological constant (or equivalently the temperature) is a free parameter, exactly like in the $\CGHS$ model.  In the next section, we will see that the effective 2d theory is actually equivalent to  the $\CGHS$ model. This will allow to define this theory using the boundary conditions and boundary action described in Sec.~\ref{The CGHS model}.

\subsubsection{Equivalence with $\CGHS$}\label{BTZeqCGHS}

We have obtained in \eqref{2dtheoryBTZ} the 2d effective theory governing the near horizon dynamics of the BTZ black hole. We show here that this theory is actually equivalent to the $\CGHS$ model \eqref{CGHShat}. A rationale for the equivalence of the two theories is as follows. After integration over their gauge fields, both theories are "minimal" versions of the CGHS model \eqref{CGHSaction} in which $\L$ is promoted to a field that becomes constant on-shell. It is thus reasonable to believe that there is a choice of boundary condition and boundary action for the BTZ near horizon theory \eqref{2dtheoryBTZ} that makes it equivalent to $\CGHS$.

Indeed, we will show that the two theories are related by a simple field redefinition and Gaussian integration. Let us recall the $\CGHS$ action:
\be
I_{\CGHS} =- {r_+\ov 8 G} \int d^2 x\sqrt{g}\le( \Phi R + 2 \Lambda- 2 \Lambda \ve^\mn \p_\mu A_\nu\ri)~.
\ee
We perform a field redefinition from $(A_\mu,\L)$ to $(a_\mu,X)$ given by
\be\label{fieldredefBTZ}
A = {1\/2 X} a,\qq \L =X^2+{1\/\l_3^2}~,
\ee
where $a = a_\mu dx^\mu$ is a 2d gauge field and $X$ is a scalar field. This leads to the action
\be\label{eqBTZactionX}
I_{\CGHS} =- {r_+\ov 8 G} \int d^2 x\sqrt{g}\le( \Phi R + 2 X^2 - 2 X \ve^\mn \p_\mu a_\nu + {2\/\l^2_3}\ri) + I_\p^\r{IPP} ~,
\ee
where we have performed some integration by parts, resulting in a boundary term $I_\p^\r{IPP}$.
The path integral over $X$ is a Gaussian integral and can be performed exactly. This leads to
\be\label{eqBTZaction}
I_{\CGHS}=- {r_+\ov 8 G} \int d^2 x\sqrt{g}\le( \Phi R -{1\/4} f_\mn f^\mn + {2\/\l^2_3}\ri) + I_\p^\r{IPP} ~,
\ee
which is indeed the BTZ theory \eqref{2dtheoryBTZ}. This theory needs to be defined by choosing suitable boundary conditions and boundary action at the asymptotic boundary of the near horizon region. They can be chosen  so that, under the above field redefinition, the BTZ theory is equivalent to $\CGHS$ gravity.  At the end, the total action coming from the BTZ black hole can be written
\be
I=I_{\mathrm{top}}+I_{\CGHS}+I_{\partial}~,
\ee
which is the same as the full action considered in Sec.~\ref{Euclidean path integral} to study of the $\CGHS$ path integral.

\subsubsection{Thermodynamics}

We will see that $\CGHS$ captures the leading change in entropy of the black hole under  a near horizon perturbation. Under the deformation \eqref{deformationBTZ}, the entropy variation is simply given by 
\be\label{deltaSBTZ}
\d S_\r{BTZ} = {\pi \d r_+\/ 2 G}~,
\ee
which we will compare to the entropy of $\CGHS$. The relations between the Newton's constant in two and three dimensions imply that we should identify
\be
\k= {1\/8\pi G_2} =  {r_+\/4 G_3}~,
\ee
and we can read off  that $\phi_r = {1\/r_+}$ from the expansion \eqref{BTZexpansionValues}. The entropy obtained from the on-shell action of $\CGHS$ evaluates to 
\be
S_{\CGHS} = {\pi \d r_+\/2 G}~,
\ee
which matches the variation of BTZ entropy \eqref{deltaSBTZ}. 

This match actually follows from the fact that $\Phi$ captures the change of the horizon area and that the CGHS entropy is given by the value of the dilaton at the horizon:
\be
S_{\CGHS} ={\Phi_{r=r_h}\/4 G_2} ~.
\ee

\subsubsection{Second order dynamics}

As $\hat{g}$ is degenerate in the limit $\la\to0$, it is necessary to include the second order fields in \eqref{expBTZ} if we want to have a solution of Einstein equation at order $\la$. More generally, we need to consider the expansion at order $\la^{n+1}$ to have a solution of Einstein equation at order $\la^n$.

The second order fields in \eqref{expBTZ} will be sourced by the first order fields. Their dynamics is obtained by expanding the action \eqref{BTZreducedaction} to second order. This leads to
\bea\nt
I^{(2)}_{2d} \= -{r_+\/ 8G} \int_M d^2 x\sqrt{g}\le[ \g^\mn (\n_\mu\n_\nu - R_\mn) \Phi+{2 \/\l_3^2}\Phi+{1\/2}R\, \chi  -{1\/4} (3\Phi f_\mn+h_\mn )f^\mn \ri.\\
&& \hspace{4cm} \le. +{1\/2}\g_\rho^{~\rho} \le(\Phi R  - {1\/4} f_\mn f^\mn +  {2\/\l_3^2}\ri)\ri]+ \text{boundary term}~,
\eea
where $h= d b$ is the field strength associated to $b$. The equations for the second order fields are obtained by varying this action with respect to $\Phi,a^\mu$ and $g_\mn$.

We report here the special solution corresponding to the deformation of the BTZ geometry to second order:
\bea
\chi \=0~,\-
\g_{ab }dx^a dx^b \= - {(r_+^2 + 3 r_-^2) \/\l_3^2 r_+^2}\,r^2 du^2~, \-
b\={r_-\/\l_3} \le( -{3 r^2\/r_+^2} + {2r\/r_+}\le({\d r_+\/r_+}-{\d r_-\/r_-} \ri) + {\d r_+\d r_-\/r_+ r_-}\ri)du~.\-
\eea
We see that the 2d geometry, corrected by $\g_{ab}$ at this order,  is not flat anymore.  It would be interesting to study the second order theory in more details but we don't expect it to be universal.

\subsection{Reissner-Nordström black hole}

We now turn our attention to the Reissner-Nordström black hole in four dimensions. This is a solution of 4d Einstein gravity coupled to a $\r{U}(1)$ Maxwell field, described by the action
\be
I = -{1\/16\pi G_4} \int d^4 x \sqrt{-g}\,R - {1\/4} \int d^4 x \,\sqrt{-g}\,F_\mn F^\mn~.
\ee
The black hole solution is given by
\be\label{RNgeometry}
ds^2 = -\le(  1 - {2 G_4 M\/r} + {G_4 Q^2\/4\pi r^2}\ri) du^2 -2 dudr + r^2 d\Om^2~,\qq A = {Q\/4\pi r}du~,
\ee
written in Eddington-Finkelstein coordinates. We would like to describe perturbations of this black hole by performing a dimensional reduction on the sphere. Following \cite{Iliesiu:2020qvm}, we write the 4d metric in the following way 
\be\label{4dansatz}
ds^2 = {r_+\/\hat\Phi}\, \hat{g}_\mn dx^\mu dx^\nu + \hat\Phi^2 \,\g_{mn}(dy^m +\hat{B}^\mu \xi_\mu^m)(dy^n + \hat{B}^\nu \xi_\nu^n)~, \qq A = {1\/\sqrt{4\pi}} \hat{a}~,
\ee
where $\hat{g}_\mn$, $\hat\Phi$, $\hat{B}$ and $\hat{a}$ are 2d quantities depending only on the 2d coordinates $(t,r)$. Here, $\g_{mn}$ is the round metric on the sphere and $\hat{B}$ is an $\r{SO}(3)$ gauge field capturing diffeomorphisms of the transverse sphere. The effective 2d action obtained by dimensional reduction takes the form \cite{Iliesiu:2020qvm}
\be\label{RN2daction}
I = -{1\/4G_4} \int d^2 x\sqrt{g} \le(\hat\Phi^2 R+ 2 r_+\hat\Phi^{-1}+ {1\/3 r_+} \hat\Phi^{5}\, \Tr(\hat{H}_\mn \hat{H}^\mn) \ri) - {1\/4 r_+} \int d^2 x \sqrt{\hat{g}}\,\hat\Phi^{3} \hat{f}_\mn \hat{f}^\mn~,
\ee
where we have defined $\hat{H} = d \hat{B}$ and $\hat{f}= d\hat{a}$. 

We would like to study this action for near horizon perturbations. As in the previous section, we first apply to \eqref{RNgeometry} the near horizon limit 
\be
r\ra r_+ +\la r~,
\ee
with no redefinition of $t$, together with a deformation of the parameters{}
\be\label{defRN}
r_+\ra r_+ +\la\d r_+,\qq r_-\ra  r_--\la \d r_-~.
\ee
Writing the resulting metric in the form \eqref{4dansatz}, this leads to a near horizon expansion of the fields 
\begin{align}\label{RNexpansion}
\hat{g} &= \la g+O(\la^2)~, & \hat\Phi &= r_+( 1+\la\Phi) + O(\la^2)~,\-
\hat{a} &= {Q\/4\pi r_+}du + {\la\/r_+} a + O(\la^2)~,&\hat{B} &= {\la\/r_+} B + O(\la^2)~.
\end{align}
with the following values
\be\label{defRNvalues}
g = 2(P_0r + T_0)du^2 -2  du dr,\qquad \Phi = {r\/r_+},\qquad a= - {Q\/4\pi r_+} r du ~, \qquad B=0
\ee
and with $P_0 = -{1\/2r_+^2}(r_+-r_-)$ and $ T_0 = {1 \/ 2 r_+^2}(r_+-r_-) \d r_+$. This shows that the near horizon geometry is two-dimensional Rindler space at the  temperature of the black hole. The absence of decoupling limit is reflected by the fact that the strict limit $\la\to 0$ is degenerate. 

To describe more general perturbations, we use the the same expansion \eqref{RNexpansion} but allowing $g,\Phi,a$ and $B$ to be arbitrary functions of the 2d coordinates. Expanding the action  \eqref{RN2daction} gives
\bea
I_\r{2d} \= I_\r{2d}^{(0)}+\la I_\r{2d}^{(1)}+ O(\la^2)~.
\eea
where we have
\bea
I_\r{2d}^{(0)} \= -{r_+^2\/4G_4}\int_M d^2x \sqrt{g}\, R+  \text{boundary term}\-
I_\r{2d}^{(1)} \=  -{r_+^2\/2G_4}\int_M d^2x \sqrt{g}\le(\Phi R + {1\/6}\Tr(H_\mn H^\mn)+{1\/r_+^2}\ri) - {1\/4} \int_M d^2 x \sqrt{g} f_\mn f^\mn\\
&& \hspace{9cm} +  \text{boundary term}~,
\eea
and we have defined $H=d B$ and $f=da$. 

As in the BTZ discussion, boundary terms need to be specified at an imaginary boundary close to the horizon, interpreted as the asymptotic boundary of the near horizon theory. The  zeroth order term is again the topological term $I_\r{top}= -S_0 \chi$ with $S_0 = {\pi r_+^2/  G_4}$ is the entropy of the unperturbed black hole. At first order, we obtain CGHS gravity with matter consisting of a $\r{U}(1)$ gauge field $a_\mu$ and an $\r{SO}(3)$ gauge field $B_\mu$. After putting the matter on-shell, we obtain the CGHS model  \eqref{CGHSaction} with 
\be
\L = {1\/12}\Tr(H_\mn H^\mn)+{1\/2r_+^2} + {G_4 \/4r_+^2}f_\mn f^\mn~,
\ee
which is constant on-shell. It can be checked that with the values \eqref{defRNvalues}, this expression is proportional to the black hole temperature as expected from \eqref{relbetaLambda}.

This shows that CGHS gravity governs the near horizon dynamics of this black hole. Using a procedure similar to Sec.~\ref{BTZeqCGHS}, it can be further shown that this theory is equivalent to $\CGHS$ (with additional gauge fields). This allows us to use the boundary conditions and boundary action of $\CGHS$ to define the near horizon theory. 

The main result here is the existence of a $\CGHS$ sector in the near horizon theory, that we expect to be universal, and which captures the linearized change in entropy due to the perturbation. Indeed, the entropy of $\CGHS$ associated to the solution \eqref{defRNvalues} is
\be
S_{\CGHS} ={\Phi|_{r=r_h}\/4 G_2} = {2\pi r_+\d r_+\/G_4}~,
\ee
and matches the change of black hole entropy under the deformation  \eqref{defRN}.

\subsection{$\CGHS$ universality and the first law}

The $\CGHS$ model appears to describe a universal sector of the near horizon perturbations of non-extremal black holes, as we have seen for the BTZ and Reissner-Nordström geometries.

A general argument for this could be made along the lines of effective field theory (EFT). The point is that as we go near the horizon of a generic black hole, we expect the dynamics to be described by a two-dimensional theory involving the time and the radial direction. The CGHS Lagrangian
\be
I ={\k\/2} \int d^2 x\sqrt{g}\,(\Phi R+2\L)+\dots
\ee
is the lowest expression in terms of number of derivatives that is linear in $\Phi$, the scalar field controlling the area of the horizon, and which has two-dimensional Rindler spacetime as a solution. Since the temperature can be changed within the same theory, we should actually obtained something equivalent, or closely related, to the $\CGHS$ model, as we have shown in the above examples.

The universal appearance of CGHS gravity provides an EFT-inspired argument for the first law. The point is that the $\CGHS$ entropy satisfies the area law as given by the value of the dilaton at the horizon
\be\label{firstlawCGHS}
\d S = S_{\CGHS} = {\Phi|_{r=r_h}\/ 4 G_2}~,
\ee
where $\d S$ is the entropy associated with the perturbation. This shows that in higher dimensions, the entropy is proportional to the change in area of the black hole horizon, and provides a rather general argument for the first law of black hole thermodynamics.

Note that the first law becomes trivial if we define the black hole entropy as a Euclidean path integral. Using only Hamiltonian methods, the general proof of the first law is a celebrated result \cite{Wald:1993nt, Iyer:1994ys}. Since the formula \eqref{firstlawCGHS} for the entropy of $\CGHS$  can be derived purely by Hamiltonian methods \cite{Afshar:2019axx}, our argument for the first law can be made purely Hamiltonian. It also gives a kind of EFT understanding: we can view the universality of the first law as a consequence of the universal $\CGHS$ sector near the horizon. 

\section{Dual ensemble and baby universes}\label{Dual ensemble and baby universes}

This section is dedicated to the construction and study of the ensemble dual to $\CGHS$.  The latter turns out to be a simple a Gaussian ensemble. We study the corresponding baby universe Hilbert space and  the $\alpha$-states. We describe how factorization is achieved on the gravity side. In particular we derive a ``wormhole = diagonal'' identity, similar to the one discussed in \cite{Blommaert:2019wfy,Blommaert:2020seb,SaadTalk} for JT gravity. We then study the annealed and quenched free energies of the model in an effort to clarify the proposal of \cite{Engelhardt:2020qpv}. In particular, we show that the quenched free energy, which can be computed directly here, is non-monotonous. We comment on the interpretation of this result.

The dual ensemble suffers from the apparent issue that some members can have a negative density of states. We show that this can be cured by non-perturbative effects with respect to the topological expansion. We also comment on a matrix model completion of $\CGHS$.

\subsection{The Gaussian ensemble}
\label{The Gaussian ensemble}

We would like to understand which ensemble reproduces the correlation functions obtained in Sec.~\ref{Euclidean path integral}. The variable $Z(\beta)$ is Gaussian because its connected three and higher-point functions vanish. Instead of working with $Z(\b)$, it is more convenient to work with the (unnormalized) density of states $\rho(E)$ defined by the relation
\be
 Z(\b) = \int_0^\infty dE \, \rho(E) e^{-\b E}~.
\ee
The density of states has the following connected correlators
\be\label{pointfunctions}
\ln\rho(E)\rn  =  2\pi e^{S_0}E\,,\qq \ln \rho(E_1) \rho(E_2)\rn_c  = 4\pi^2\d(E_1-E_2)~,\qq \ln \rho(E_1) \rho(E_2)\rho(E_3)\dots\rn_c  = 0~,
\ee
and is also a Gaussian variable. We have absorbed here the constant term $2\pi\kappa\phi_h$ in a redefinition of $S_0$ and have set $\gamma=1$. The advantage of working with $\rho(E)$ is that the above two-point function implies that $\rho(E_1)$ and $\rho(E_2)$ are independent variables for $E_1\neq E_2$.

We can view these correlation functions as that of a one-dimensional field theory of the variable $\rho(E)$ with action
\begin{equation}
S[\rho] ={1\/8\pi^2}\int_0^\infty dE(\rho(E)-2\pi e^{S_0} E)^2 ~.
\label{Gaussianweight}
\end{equation}
The generating functional of this ensemble is
\be
\mathcal{Z}[J] = \int D\rho \exp\le( -S[\rho]+\int_0^\infty  dE \,\rho(E) J(E)\ri)~.
\label{GaussianZ}
\ee
Taking the logarithm, we can generate all the connected correlation functions of $\rho(E)$  using
\be
\ln \rho(E_1)\rho(E_2)\dots\rho(E_n)\rn_c=\le. {\d\/\d J(E_1)}{\d\/\d J(E_2)}\dots {\d\/\d J(E_n)}\log \mathcal{Z}[J]\ri|_{J=0}~.
\ee
We can perform the Gaussian integral exactly and we find
\be
\log \mathcal{Z}[J] =   \int_0^\infty dE \le(2 \pi^2 J(E)^2 + 2\pi e^{S_0} E J(E)\ri)+\mathrm{const}~.
\ee
One can easily check that this reproduces the connected correlation functions of $\CGHS$. We are simply dealing with infinite-dimensional Gaussian integrals in the density of states. The more surprising (and actually worrying) fact is that, to reproduce the correlation functions of $\CGHS$, we need to integrate over densities of states that can take negative values: $\rho(E)<0$. This suggests that some members of the ensemble are not physical, we will come back to this in Sec.~\ref{Truncated Gaussian ensemble} and show that this problem can be cured by non-perturbative corrections.

Moreover, from the connected two-point function we can deduce that there is no correlation between the density of states at two different energies. In other words, there is no eigenvalue repulsion (beyond the ultra-local one), which is in tension with the expectation that gravitational theories are dual to chaotic systems. Another way to diagnose the eigenvalue repulsion is to consider the spectral form factor, given by the following analytic continuation 
\begin{equation}
\langle Z(\beta+it)Z(\beta-i t)\rangle=4\pi^2\left[\frac{e^{2S_0}}{(\beta^2+t^2)^2}+\frac{1}{2\beta}\right]~.
\end{equation}
As $t$ increases, the spectral form factor exhibits a plateau, which reflects the discreteness of the spectrum, but no ramp, characteristic of eigenvalue repulsion. This ultra-locality of the eigenvalue repulsion is also present in three dimensions, in the low-temperature limit of the two-point function when averaging over the Narain ensemble \cite{Cotler:2020hgz,Cotler:2020ugk}. This has to be contrasted with the spectral form factor of chaotic systems \cite{Cotler:2016fpe}.

\begin{figure}
  \centering
  \includegraphics[width=11cm]{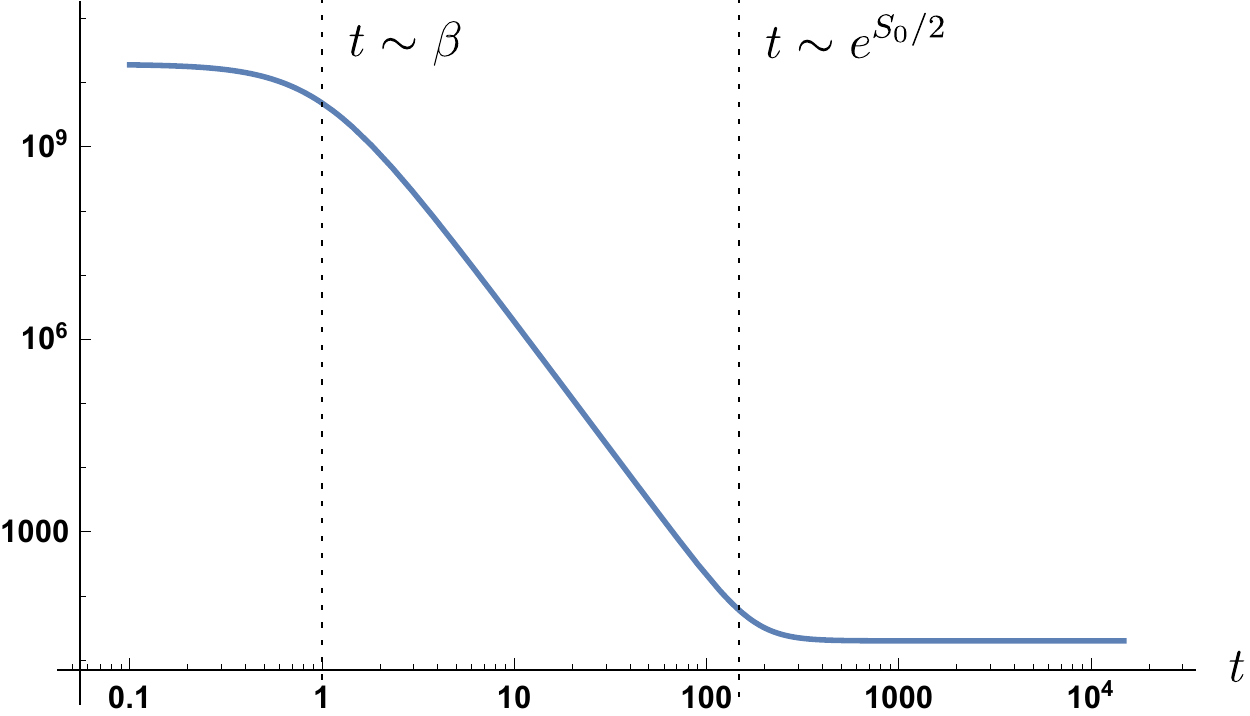}
  \caption{Spectral form factor of $\CGHS$. Plot made with $S_0=10$.}\label{FigSFF}
\end{figure}
 The likely reason for the lack of eigenvalue repulsion is that we are working in the wrong thermodynamical ensemble. In order to really probe eigenvalue repulsion, one needs to fix all the charges of the system, because we are otherwise superposing the spectrum of independent sectors which should always give Poisson statistics.  The corresponding charge should be the one associated to the $\mathrm{U}(1)$ gauge field. The quantities we are computing in \eqref{pointfunctions} would actually correspond to correlation functions of $\rho(E,\mu=0)$ where $\mu$ is the chemical potential associated to the $\mathrm{U}(1)$ charge. Then the right quantity to consider should be the Fourier transform $ \rho(E,Q)=\int d\mu\, \rho(E,\mu)e^{i\mu Q}$ which corresponds to the density of states at fixed charge. This is the quantity that should exhibit eigenvalue repulsion. 

One should notice that whatever boundary conditions we impose on the gauge field, the model will remain Gaussian\footnote{Gaussian in the sense that the connected $n$-point functions of $Z(\beta,\mu)$ will still vanish for $n>2$.} since only the disk and cylinder geometries will contribute. One should bear in mind that a Gaussian model can still have non-trivial eigenvalue repulsion encoded in its two-point functions, while the negativity of the density of states for some members of the ensemble (due to the negative tail of the Gaussian) is cured by non-perturbative effects. 
We come back to this in Sec. \ref{Non-perturbative completion}.

\subsection{Baby universes and factorization}

The connection between ensemble averages and baby universes goes back to the works of Coleman, Giddings and Strominger in the 80's \cite{Giddings:1987cg, giddins1988, GIDDINGS1989481, coleman1988}. This connection was revived recently in the context of holography where the ensemble is seen as an ensemble of boundary theories \cite{Marolf:2020xie}. In the baby universe picture, the non-factorization is due to the fact that we have picked a particular baby universe state to compute boundary correlators: the Hartle--Hawking or no-boundary state. Factorization occurs when computing correlators in the so-called $\alpha$-states, defined as eigenstates of the operator that inserts a boundary. In other words, $\alpha$-states give access to one member of the ensemble. We would like to understand the baby universe Hilbert space of $\CGHS$, the corresponding $\alpha$-states and the bulk interpretation of factorization.

\paragraph{Baby universe Hilbert space.}
The nice thing about our model is its simplicity. This simplicity allows for a complete characterization of the baby universe Hilbert space. We can interpret the gravitational path integral $\langle Z(\beta)\rangle$ as the one-point function of an operator $\hat{Z}(\beta)$ in the no-boundary state $\ket{\mathrm{HH}}$:
\begin{equation}
\langle Z(\beta)\rangle=\bra{\mathrm{HH}}\hat{Z}(\beta)\ket{\mathrm{HH}},
\end{equation}
while an $n$-point function of this operator will correspond to the gravitational path integral with $n$ boundaries. The operator $\hat{Z}(\beta)$ is understood as an operator that creates a boundary on the no-boundary state:
\begin{equation}
\hat{Z}(\beta)\ket{\mathrm{HH}}=\ket{Z(\beta)},
\end{equation}
and acting $n$ times with this operator creates $n$ thermal boundaries.

The structure of the $n$-point function of the operator $\hat{Z}(\beta)$, or equivalently $\hat{\rho}(E)$, in the no-boundary state is that of a free field in the vacuum and we can implement this structure using creation and annihilation operators: 
\begin{equation}
\hat{\rho}(E)=2\pi (e^{S_0}E+a_E+a^\dagger_E)\quad \text{with} \quad [a_E, a^\dagger_{E'}]=\delta(E-E').
\end{equation}
This ensures that all $n$-point connected correlation functions are vanishing for $n\geq 3$ and gives the right values to the one-point and two-point functions. We conclude that the baby universe Hilbert space is a Fock space whose vacuum is given by the no-boundary state, \emph{i.e.} is annihilated by all the $a_E$'s. All the states are obtained by acting with creation operators on $\ket{\mathrm{HH}}$. The third quantized theory is nothing but a collection of oscillators labeled by the boundary energy $E$ and the density of states operator $\hat{\rho}(E)$ behaves like a position operator. The same Fock space appears in \cite{Marolf:2020xie} in the context of baby universe perturbation theory.

\paragraph{\pmb{$\alpha$}-states.} We would like to determine the eigenstates of the density of states operator, the so-called $\alpha$-states, to understand better factorization. Since $\hat{\rho}(E)$ is simply the usual position operator of a harmonic oscillator, the $\alpha$-states are simply
\begin{equation}
\ket{\alpha}=\mathcal{N}\exp\left[-\frac{1}{2}\int_0^\infty dE\left(a^\dagger_E-\alpha(E)\right)^2\right]\ket{\mathrm{HH}},
\end{equation}
and satisfy
\begin{equation}
\hat{\rho}(E)\ket{\alpha}=2\pi( e^{S_0}E+\alpha(E))\ket{\alpha}.
\end{equation}
The normalization constant $\mathcal{N}$ is here to ensure that $\braket{\alpha}=1$. It is using this baby universe state to compute the boundary two-point function that we will have factorization, as the equality
\begin{equation}
\begin{split}
\bra{\alpha}\hat{\rho}(E_1)\hat{\rho}(E_2)\ket{\alpha} =\bra{\alpha}\hat{\rho}(E_1)\ket{\alpha}\bra{\alpha}\hat{\rho}(E_2)\ket{\alpha}
\end{split}
\end{equation}
simply follows from the fact that $|\a\rn$ is an eigenstate of $\hat\rho(E)$. This also implies that we will have factorization for correlators of the partition function. 

Another way to understand the $\alpha$-states is that they correspond to boundary conditions in the gravitational path integral which allow to isolate one member of the dual ensemble, reconciling the bulk description and its holographic interpretation. The price to pay is that we need to compute boundary correlators in a particular background state that is not the usual no-boundary state.

We have found a representation of the $\alpha$-states in terms of the creation operator, but this does not tell us how we need to modify the boundary conditions in the gravitational path integral to compute boundary correlators in that state. To do so we need to find a representation of the $\alpha$-state using an operator whose gravity dual is known, such as the position operator $\hat{\rho}(E)$, which corresponds to a fixed energy boundary condition, defined as the Laplace transform of  a thermal boundary condition. The vacuum state of a harmonic oscillator can be represented as a sum of position eigenstates with a Gaussian weight:
\begin{equation}
\ket{\mathrm{HH}}\propto \int D\alpha\exp\left[-\frac{1}{4}\int_0^\infty dE\, \alpha(E)^2\right]\ket{\alpha}.
\end{equation} 
Also, a particular $\alpha$-state, say $\ket{\alpha^\star}$ can be obtained by acting on the vacuum using the projector
\begin{equation}
\frac{1}{\bra{\alpha^\star}\ket{\mathrm{HH}}}\ket{\alpha^\star}\bra{\alpha^\star}.
\end{equation}
We can use the integral representation of the Dirac delta function to write it as 
\begin{equation}
\frac{1}{\bra{\alpha^\star}\ket{\mathrm{HH}}}\int D\alpha\int D\Pi \exp\left[\int_0^\infty dE\, i\Pi(E)(\alpha(E)-\alpha^\star(E))\right] \dyad{\alpha}{\alpha}.
\end{equation}
Now using the fact that the operator $ a_E+a^\dagger_E$ is diagonal with eigenvalues $\alpha(E)$ we obtain
\begin{equation}
\ket{\alpha^\star}=\frac{1}{\bra{\alpha^\star}\ket{\mathrm{HH}}}\int D\Pi \exp\left[\int_0^\infty dE\, i\Pi(E)((2\pi)^{-1}\hat{\rho}(E)-e^{S_0}E-\alpha^\star(E))\right]\ket{\mathrm{HH}},
\label{alphastates}
\end{equation}
This  operator creates an $\a$-state when acting on the vacuum $\ket{\mathrm{HH}}$ and is written as a functional of $\hat{\rho}(E)$, whose bulk dual is known. This makes it possible to (formally) compute in an $\a$-state in gravity by inserting this operator. So if we wanted to compute a gravitational partition function with $n$ boundaries in an $\a$-state, we would have to include this complicated linear superposition of fixed energy boundaries in addition to our our $n$ boundaries. Many new surfaces would then contribute and conspire to lead to factorization.

\paragraph{SD-branes.} In the topological model of \cite{Marolf:2020xie} it was noted that there exists a relation between the spacetime D-branes, that correspond to objects on which spacetime can end, and $\alpha$-states. A similar relation exists in our Gaussian model. In our case, the operator that creates a spacetime D-brane, parametrized by a function $g(E)$, is 
\begin{equation}
\exp\le[\int_0^\infty g(E) \hat{\rho}(E)dE\ri]~.
\end{equation}
This operator acting on the vacuum creates a state that we denote $\ket{\mathrm{SDbrane}_g}$. Using the formula \eqref{alphastates}, we obtain that the $\alpha$-states are obtained as a Fourier transform of complex spacetime D-branes:
\begin{equation}
\ket{\alpha}\propto \int Dg \,\exp\left[ -2\pi i \int_0^\infty dE\,g(E)(e^{S_0}E+\alpha(E))\right]\ket{\mathrm{SDbrane}_{i g}}.
\end{equation}
This is yet another way to give a geometrical interpretation to the $\alpha$-state.

This is actually expected on general ground as an eigenstate of the operator $\hat{Z}$ with eigenvalue $\alpha$ can be created by acting on the no-boundary state $\ket{\mathrm{HH}}$ with the Dirac-delta operator $\delta(\hat{Z}-\alpha)$. Using the Fourier representation of the Dirac-delta gives 
\begin{equation}
\ket{\alpha}=\int dg\, e^{-ig\alpha}e^{i g \hat{Z}}\ket{\mathrm{HH}}=\int dg\, e^{-ig\alpha}\ket{\mathrm{SDbrane}_{i g}},
\end{equation}
where the $\alpha$-state is written as a sum of spacetime D-brane states. So $\alpha$-states should always admit some  spacetime D-brane representation involving a Fourier transform.

\paragraph{Factorization.} We would like to go further in our understanding of factorization in the bulk. From the dual point of view, it is clear that factorization occurs only if we are in an $\a$-state but we would like to see what it implies in gravity. For the two-point function, we will see that this implies a ``wormhole = diagonal'' identity, similar to that of JT gravity \cite{Blommaert:2019wfy,Blommaert:2020seb,SaadTalk}.

In order to show this we need to discretize the model. Instead of having a continuum of energies, we will take a set of energies $E_i$ with $i=1,\dots,L$ and use the Riemann sum formula
\begin{equation}
\int_0^\Lambda \rho(E)e^{-\beta E}dE=\lim_{L\to \infty}{\L\/L}\sum_{i=1}^L \rho(E_i)e^{-\beta E_i}~,
\end{equation}
which is valid as long as the spacings between the $E_i$ go to zero as $L\to+\infty$. We also introduce a cutoff $\L$ that is taken to infinity at the end. For example, a simple choice is to take $E_i = {i\L\/L}$ for $i=1,\dots,L$.

Instead of having a continuum of Gaussian variables, we now have a finite set of them given by the $\rho(E_i)$ for $i=1,\dots,L$. To obtain the probability distribution on these variables, we discretize the integral over $E$ in \eqref{Gaussianweight} which leads to
\begin{equation}
S[\rho]=\frac{1}{8\pi^2}\lim_{L\to \infty} \frac{\Lambda}{L}\sum_{i=1}^{L}\left(\rho(E_i)-2\pi e^{S_0}E_i\right)^2~.
\end{equation}
From this we can deduce the following one-point and two-point functions for the discretized model
\begin{equation}
\langle \rho(E_i) \rangle =2\pi e^{S_0} E_i,\qq \langle\rho( E_i)\rho( E_j) \rangle_c = 4\pi^2\frac{L}{\Lambda}\delta_{ij}~.
\label{DiscreteCorr}
\end{equation}
These correlation functions are also reproduced using a Fock space representation of the Hilbert space. We now have a discrete set of boundary operators  that can be written in terms of creation and annihilation operators
\begin{equation}
\hat{\rho}(E_i)=2\pi(e^{S_0}E_i+a_i+a^\dagger_i)~.
\end{equation}
In the discretized case, the $\alpha$-states are labeled by a collections of numbers $\alpha_i$ corresponding to the eigenvalues of the $L$ operators $\hat{\rho}(E_i)$. We can build them out of the vacuum in the following way
\begin{equation}
\ket{\alpha}=\frac{1}{\bra{\alpha}\ket{\mathrm{HH}}}\prod_{i=1}^L \mathcal{O}_{\alpha_i}(\hat{\rho}(E_i))\ket{\mathrm{HH}},
\end{equation} 
where we have defined
\be
\mathcal{O}_{\alpha_i}(\hat{\rho}(E_i))\equiv \int dp_i \exp\left[i \,p_i\left((2\pi)^{-1}\hat{\rho}(E_i)-e^{S_0}E_i-\alpha_i\right)\right]~.
\ee
We see that $|\a\rn$ is a simultaneous eigenstate of the operators $\hat{\rho}(E_i)$ with eigenvalues $2\pi(e^{S_0}E_i+\alpha_i)$. The discretization is useful to be able to write  the $\alpha$-state as the above product of operators acting on $|\r{HH}\rn$. The operator $\mathcal{O}_{\alpha_i}(\hat{\rho}(E_i))$ is responsible for fixing the eigenvalue of $\hat{\rho}(E_i)$ to be $\alpha_i$. In the limit $L\to\infty$, this product fixes the whole function $\alpha(E)$. They are the equivalent of the eigenbranes in JT gravity \cite{Blommaert:2019wfy}. 

Trivially the $\alpha$-states satisfy the following equality
\begin{equation}
\bra{\mathrm{HH}}\hat{\rho}(E_i)\hat{\rho}(E_j)\ket{\alpha}\bra{\mathrm{HH}}\ket{\alpha}=\bra{\mathrm{HH}}\hat{\rho}(E_i)\ket{\alpha}\bra{\mathrm{HH}}\hat{\rho}(E_j)\ket{\alpha}\, ,
\end{equation}
which is equivalent to 
\begin{equation}
\begin{split}
&\bra{\mathrm{HH}}\hat{\rho}(E_i)\hat{\rho}(E_j)\prod_{k=1}^L \mathcal{O}_{\alpha_k}(\hat{\rho}(E_k))\ket{\mathrm{HH}}\bra{\mathrm{HH}}\prod_{\ell=1}^L \mathcal{O}_{\alpha_\ell}(\hat{\rho}(E_\ell))\ket{\mathrm{HH}}\\&=\bra{\mathrm{HH}}\hat{\rho}(E_i)\prod_{k=1}^L \mathcal{O}_{\alpha_k}(\hat{\rho}(E_k))\ket{\mathrm{HH}}\bra{\mathrm{HH}}\hat{\rho}(E_j)\prod_{\ell=1}^L \mathcal{O}_{\alpha_\ell}(\hat{\rho}(E_\ell))\ket{\mathrm{HH}}\,.
\end{split}
\end{equation}
This equality is trivially obtained from the fact that $\ket{\alpha}$ is an eigenstate of the density operator. However, it becomes non-trivial when we consider its gravitational interpretation, and provides an interesting interpretation of the wormhole contribution. 

\begin{figure}
  \centering
  \includegraphics[width=13cm]{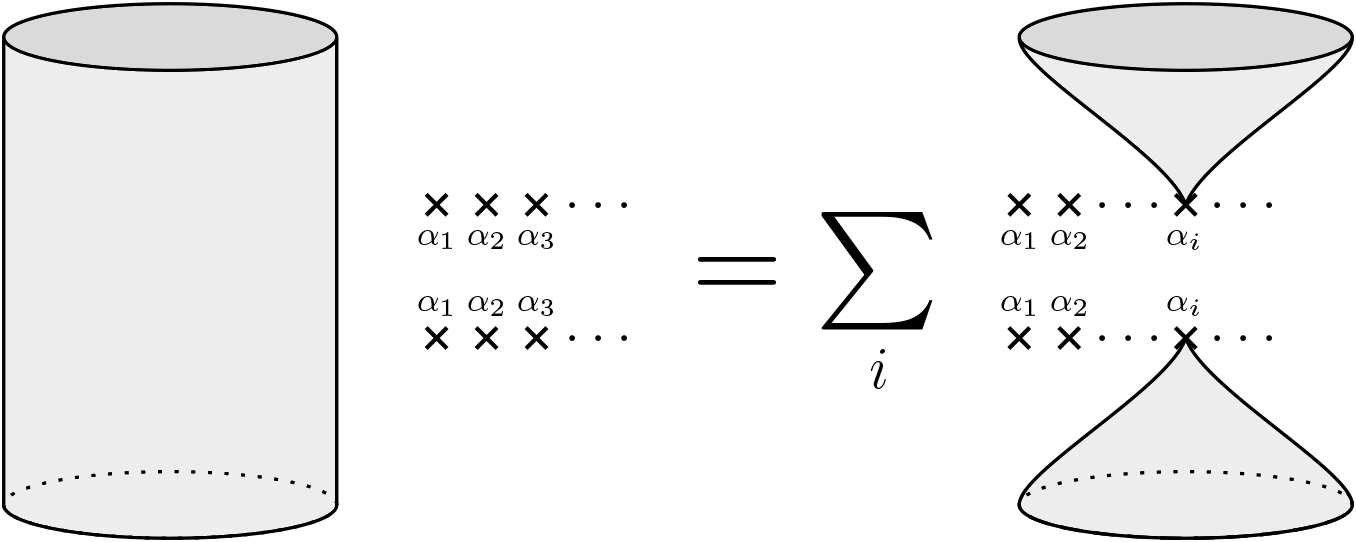}
  \caption{$\CGHS$ version of the ``wormhole = diagonal'' identity which follows from the factorization of the two-point function in an $\a$-state. A cross $\a_i$ corresponds to the insertion of the operator $\cO_{\a_i}$ and can be viewed as a coherent superposition of fixed energy boundaries. There is also a sum over all cylinders connecting every pair of free crosses in each line (but not crosses on different lines) which we have not represented here.}\label{Fig:WHdiag}
\end{figure}

Each operator $\mathcal{O}_{\alpha_i}(\hat{\rho}(E_i))$ is a functional of $\hat{\rho}(E_i)$ and can be (formally) interpreted as a coherent superposition of fixed energy boundaries. We represent this boundary condition with a cross labeled by the eigenvalue $\alpha_i$ that it fixes. On both sides of the equality, the boundary condition consists of two fixed-energy boundaries and two copies of $L$ crosses. The only thing that changes between the two sides is the type of surfaces that are allowed to contribute. In particular, surfaces that connect the two fixed-energy boundaries are allowed only on the left side while surfaces that connect both fixed-energy boundaries to the \emph{same} eigenvalue are allowed only on the right side. Cancelling the contributions that are present on both sides, we end up with an interesting gravitational identity  represented in  Fig.~\ref{Fig:WHdiag}. This equality is similar to the ``wormhole = diagonal'' obtained for JT gravity \cite{Blommaert:2019wfy,Blommaert:2020seb,SaadTalk}.

\subsection{Quenched versus annealed free energy}
\label{Quenched versus annealed free energy}

The ensemble interpretation of the boundary theory comes with a modification of the natural way to compute certain quantities. Indeed if one computes a certain averaged quantity $\langle Z\rangle$ in gravity and wants to know the value of $\langle f(Z)\rangle$, one cannot simply apply $f$ to the result of $\langle Z\rangle$. This is simply because 
\begin{equation}
\langle f(Z)\rangle\neq f(\langle Z\rangle).
\end{equation}
An example of this issue was studied in \cite{Engelhardt:2020qpv} and corresponds to the computation the free energy. Indeed the usual way to compute the free energy in gravity is to compute 
\begin{equation}
{}-T \log Z(\b)\,.
\label{wrongF}
\end{equation}
in terms of the gravitational partition function at temperature $T$. 
However, if the gravitational path integral computes averaged quantities, this formula does not give the averaged free energy. This ``wrong'' quantity corresponds to the \emph{annealed} free energy while the ``right'' one should be the \emph{quenched} free energy:
\begin{equation}
F_{\mathrm{ann}}=-T  \log\langle Z(\beta)\rangle~,\qq F_{\mathrm{quen}}=-T \langle \log Z(\beta)\rangle~,
\end{equation}
which are in general different.  The annealed free energy  contains the contribution of all surfaces with one boundary. The quenched free energy corresponds to an even stronger modification where the geometrical interpretation is less clear.

In the $\CGHS$ model, the annealed free energy is given by
\begin{equation}
F_{\mathrm{ann}}=-T\log\left(2\pi e^{S_0}T^2\right).
\label{Fann}
\end{equation}
This function has a local maximum at temperature $T_{\mathrm{max}}=e^{-S_0/2}/(\sqrt{2\pi}e)$. This is an unphysical feature because it leads to a negative thermodynamical entropy $S=-\partial F/\partial T$ for $T<T_{\mathrm{max}}$. One thing to mention also is that the gravitational description is expected to break down at $T\lesssim e^{-S_0}$ at which point the discreteness of the spectrum becomes important. Nevertheless, the thermodynamical entropy is negative in a range of temperature $e^{-S_0}\ll T \ll e^{-S_0/2}$ where we can trust the gravitational description.

One of the proposal in \cite{Engelhardt:2020qpv} is that going to the quenched free energy, which is the quantity of real interest, might cure this non-monotonicity problem. To compute the quenched free energy, they propose the replica trick
\begin{equation}
 \langle \log Z(\beta)\rangle=\lim_{n\to 0}\frac{ 1}{n}(\langle  Z(\beta)^n\rangle-1)~.
\end{equation}
In gravity, the computation of the quenched free energy will then involve wormholes connecting the various replicated boundaries. Comparing the fully disconnected contribution and the one where each boundary is connected to another one by a cylinder, they estimate that the quenching should become relevant at temperature lower than a critical one $T_c=2^{-1/3}e^{-2S_0/3}$.\footnote{We have reintroduced the constant $S_0$ in their result.} One should notice that for small $S_0$\footnote{Precisely for $S_0<6+\log 2\pi^3$.} the critical temperature is larger than $T_{\mathrm{max}}$, which means that in principle wormholes could cure the non-monotonicity of the free energy. However, the analysis of \cite{Engelhardt:2020qpv} is inconclusive because the analytic continuation in $n$ is ambiguous. When $S_0$ is large, the critical temperature $T_c$ becomes smaller than $T_{\mathrm{max}}$ and there is no hope that replica wormholes would correct the behavior of the free energy in the range $T_c < T< T_{\mathrm{max}}$.  

Here, we use  another method to compute the quenched free energy of the $\CGHS$ model. In fact, it can be computed directly and unambiguously using the dual ensemble. To do so we notice that the free energy is computed by considering the variable $Z(\beta)$ at fixed temperature. The one-point and connected two-point functions for this variable are
\begin{equation}
\langle Z(\beta) \rangle = 2\pi e^{S_0}/\beta^2 ,\qq \langle Z(\beta)^2 \rangle_c = 2\pi^2/\beta\, ,
\label{correlZ}
\end{equation}
and all higher connected moments  vanish. This means that at fixed $\beta$, we can take the probability distribution of the variable $Z(\beta)$ to be Gaussian. In other words, we have for any functional $\cF$:
\begin{equation}
\ln \cF(Z(\b)\rn ={\sqrt{\b}\/2\pi^{3/2}}\int_{-\infty}^{\infty} dZ \exp\left(-\frac{\beta}{4\pi^2}\left(Z-2\pi e^{S_0}/\beta^2\right)^2\right) \cF(Z)~.
\label{1d_ensemble}
\end{equation}
Note that in the $\CGHS$ ensemble, it is not true that $Z(\b)$ is a Gaussian variable since it has non-trivial correlations with $Z(\b')$ for $\b'\neq\b$. The precise statement is that the Gaussian is the marginal probability distribution of $Z(\b)$, defined after integrating out $Z(\b')$ for all $\b'\neq\b$ in the ensemble.

We define the quenched free energy of $\CGHS$ to be
\begin{equation}
F_{\mathrm{quen}}\equiv \Re{\left[-T\langle\log Z(\beta)\rangle\right]}=-T\langle\log \abs{Z(\beta)}\rangle,
\end{equation}
where the real part is necessary because $Z(\b)$ can become negative in the ensemble. This is the same prescription that was used in \cite{Engelhardt:2020qpv} and can be interpreted as removing a pure phase in the partition function.\footnote{The negativity of $Z(\b)$ is due to the negativity of $\rho(E)$ in the $\CGHS$ ensemble. This can be cured by non-perturbative corrections as shown in the next section. We don't expect these effects to be able to resolve the non-monotonicity of the free energy since they only appear at temperatures $T \lesssim e^{-S_0}$. For this reason, it is enough to analyze this issue in $\CGHS$. }

The integral over $Z$ can be done exactly and leads to the quenched free energy
\begin{equation}\label{diffF}
F_{\mathrm{quen}}=F_{\mathrm{ann}}+\frac{T}{2}\le(\gamma+\log (4 e^{2S_0} T^3)+\le.  {d\/da}{}_1F_1\left(a,\tfrac12,-e^{2S_0}T^3\right)\right|_{a=0}\ri),
\end{equation}
where $F_{\mathrm{ann}}$ is given in \eqref{Fann} and $\gamma$ is the Euler constant. This is the exact quenched free energy of the $\CGHS$ model. This result does not suffer from any ambiguity as it is computed directly from the ensemble.

  \begin{figure}
\begin{center}
  \begin{tabular}{cc}
    \subf{\includegraphics[width=6.8cm]{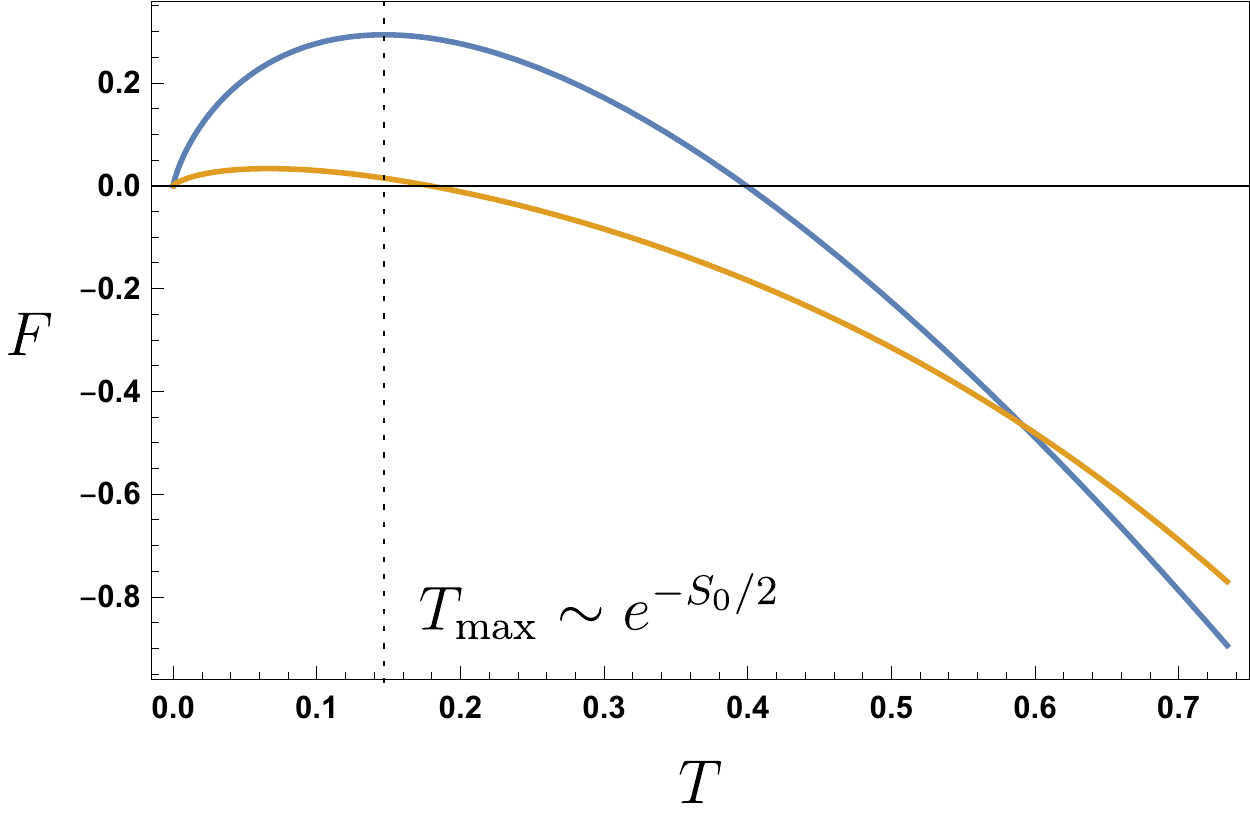}}{\hspace{0.6cm}$S_0=0$} &
    \subf{\includegraphics[width=9.6cm]{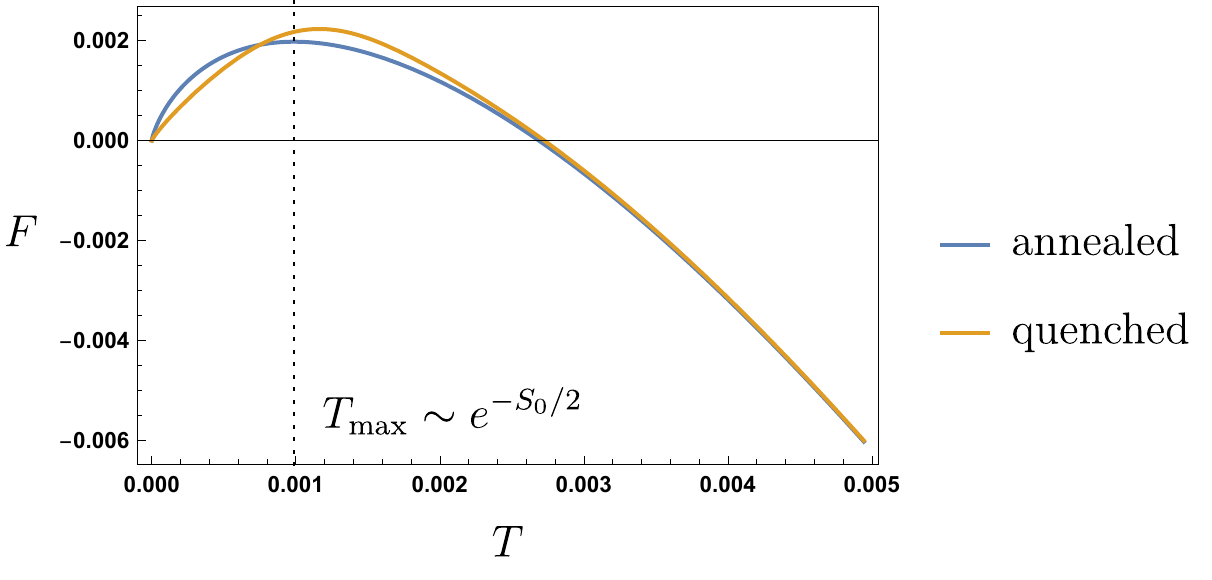}}{\hspace{-1.6cm}$S_0=10$} 
  \end{tabular}
\end{center}
\caption{Quenched and annealed free energies for the $\CGHS$ model. Both free energies have a bump around $T\sim e^{-S_0/2}$ and become almost identical at large $S_0$.}\label{Fig:plotQA}
\end{figure}

\begin{figure}
  \centering
  \includegraphics[width=13cm]{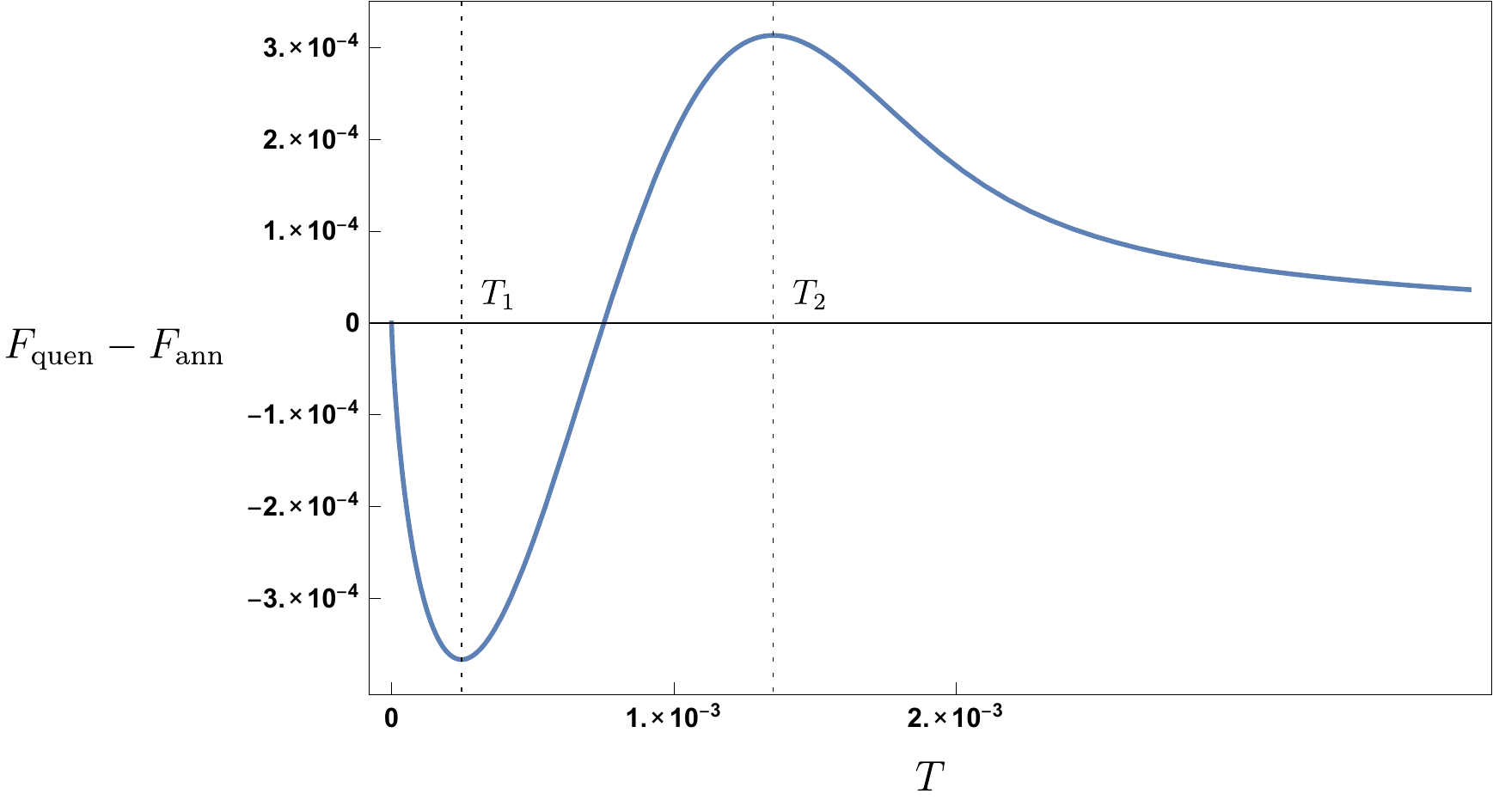}

  \caption{Difference of the quenched and annealed free energies in $\CGHS$. The temperatures corresponding to extrema are $T_1 = \a_1 e^{-2S_0/3}$ and $T_2= \a_2 e^{-2S_0/3}$ where $\a_1\approx 0.195,\a_2\approx 1.06 $  are solutions of a transcendental equation independent of $S_0$. This shows that the correction due to replica wormholes becomes only appreciable for $T \lesssim e^{-2S_0/3}$, confirming the naive estimate. The plot was made for $S_0=10$.}\label{FigDiff}
\end{figure}

We find that the quenched free energy is still problematic since it displays a bump which is qualitatively the same as the one in the annealed free energy. At small $S_0$ the bump is a little bit displaced but at large $S_0$, the quenched and annealed free energies are almost exactly the same, see Fig.~\ref{Fig:plotQA}. This was expected as wormholes should contribute at temperatures of order $T\sim e^{-2 S_0/3}$ which corresponds to a region far to the left of the bump. We can also check this by studying the difference between quenched and annealed. Indeed, we find in Fig.~\ref{FigDiff} that the difference becomes important at $T\sim e^{-2 S_0/3}$. This confirms the naive estimate for $T_c$ obtained  using the wormhole contribution to the replica trick formulation. We see that even in the region where the wormholes contribute the sign of the thermodynamical entropy is not corrected and remains negative.

\paragraph{Discrete versus continuous.} We would like to end this section with a comment on the distinction between discrete and continuous systems and its implication for the free energy. Any well-defined discrete system admits a monotonous free energy.\footnote{We prove this statement in a footnote of App.~\ref{From discrete to continuous}.} Therefore, any ensemble of discrete systems will have a monotonous quenched free energy. Now if a discrete system admits a continuous limit, nothing ensures that the resulting free energy will be monotonous. In App.~\ref{From discrete to continuous} we give an example of discrete system whose continuous limit gives the averaged partition function of the $\CGHS$ model, so the free energy becomes non-monotonous in the continuous limit.\footnote{Another example of continuous system that has a non-monotonic free energy is the free non-relativistic particle whose energy density goes like $\sqrt{E}$. The corresponding free energy is $F=-T \log\left(\frac{1}{2}\sqrt{\pi}\,T^{3/2}\right)$ and is non-monotonous.}

The Euclidean gravitational path integral admits an ensemble interpretation, so a reasonable question to ask is whether the members of this ensemble have a monotonous free energy. In principle some of these members could be continuous systems with non-monotonous free energies, and if they dominate the ensemble, the resulting quenched free energy could be non-monotonous. This is actually what happens for the $\CGHS$ model, where the dominant configurations have $\rho(E)\sim 2\pi e^{S_0}E$ resulting in a non-monotonous quenched free energy.  This suggests that the quenched free energy of JT gravity might be non-monotonous due to the dominance of continuous members in the ensemble. It would be interesting to investigate if this is the case or not.

\subsection{Truncated Gaussian ensemble}
\label{Truncated Gaussian ensemble}

The $\CGHS$ ensemble seems unphysical because it has members with negative density of states.  The problem comes from the fact that the variable $\rho(E)$ is a Gaussian variable, which has support on the entire real axis. The fact that, perturbatively, $\CGHS$ does not define a proper ensemble was also noticed in \cite{Janssen:2021stl}.

In fact, we will see that this issue can be cured using non-perturbative corrections in the genus expansion (which correspond to doubly non-perturbative effects in Newton's constant). To illustrate this, we consider a simple model that has non-negative densities of states and whose correlation functions are exactly the same as the $\CGHS$ ones up to non-perturbative corrections.

Instead of the Gaussian ensemble described earlier \eqref{GaussianZ}, we consider a truncated Gaussian ensemble where we integrate only over non-negative density of states:
\begin{equation}
\mathcal{Z}^{\mathrm{TG}}[J]=\cN \int_{\rho\geq 0} D\rho\, \exp \left(-S[\rho]+\int_0^\infty dE J(E)\rho(E) \right),
\label{ZTG}
\end{equation}
where the superscript $\mathrm{TG}$ stands for truncated Gaussian and $\cN$ is a normalization factor. This model is very close to the Gaussian one as the mean value of $\rho(E)$ is proportional to $e^{S_0}$, so the truncation only removes a small tail of the probability distribution at large $S_0$.

We can compute its connected generating function exactly:
\begin{equation}
\log \mathcal{Z}^{\mathrm{TG}}[J]=\int dE \left(2 \pi^2 J(E)^2 + 2\pi e^{S_0} E J(E)+\log\left(1+\mathrm{erf}\left[\frac{2 J(E)\pi+ e^{S_0} E }{\sqrt{2}}\right]\right)\right)+\mathrm{const}~.
\end{equation}
The first two terms correspond to the Gaussian connected generating function but we have a new term that comes as a correction. As $S_0\to \infty$, the error function $\r{erf}$ goes to $1$ so we recover the connected generating function of the Gaussian distribution.

As an example we can compute the one-point function in this model
\begin{equation}
\langle \rho(E)\rangle^{\mathrm{TG}}= 2\pi e^{S_0} E+\frac{2\sqrt{2\pi} }{1+\mathrm{erf}\left( \frac{e^{S_0} E}{\sqrt{2}} \right)}e^{-e^{2S_0}E^2/2}~.
\label{TG1point}
\end{equation}
We see that the correction term is only important at energy $E \lesssim e^{-S_0}$ where we expect the  gravity description to be invalid. The correction is non-perturbative in the genus expansion (doubly non-perturbative in the Newton constant). One can check that the structure is the same for higher-point functions
\begin{equation}
\begin{split}
&\langle \rho(E_1)\rho(E_2)\rangle^{\mathrm{TG}}_c =4\pi^2 \delta(E_1-E_2)(1+ \text{non-perturbative})\, ,\\
&\langle \rho(E_1)\ldots\rho(E_n)\rangle^{\mathrm{TG}}_c =\delta(E_1-E_2)\ldots \delta(E_{n-1}-E_n)(0 +\text{non-perturbative})\, , \quad n>2\, .
\end{split}
\label{TGnpoint}
\end{equation}
The Gaussian connected $n$-point functions all receive non-perturbative corrections. In the bulk, it might be possible to view these corrections as some kind of spacetime D-branes. 

This toy model also allows to understand an interesting phenomenon which is the reduction of the baby universe Hilbert space when higher-point correlators are non-zero. 
Forgetting the non-perturbative contributions,  the truncated Gaussian model reduces to the Gaussian one. With the non-perturbative contributions, one can check that the baby universe Hilbert space reduces drastically. The way to see this is actually pretty simple. The ensemble and the basis of $\alpha$-states are related in a precise way \cite{Marolf:2020xie}:
\begin{equation}
\bra{\mathrm{HH}}\hat{\rho}(E)\ket{\mathrm{HH}}=\sum_{\alpha}\abs{ \bra{\alpha}\ket{\mathrm{HH}}}^2\bra{\alpha} \hat{\rho}(E)\ket{\alpha}
\end{equation}
which follows from  decomposing the no-boundary state into $\alpha$-states. The ensemble is just the space of $\alpha$-states with the probability distribution $p_\alpha=\abs{ \bra{\alpha}\ket{\mathrm{HH}}}^2$. In the truncated Gaussian, it corresponds to the set of functions $\rho(E)$ that are non-negative. Therefore, in this model, the set of $\alpha$-states is described by the eigenstates of $\hat{\rho}(E)$ with non-negative eigenvalues:\footnote{One could imagine that there exists another ensemble,  with a smaller amount of members, which reproduces the correlations \eqref{TG1point} and \eqref{TGnpoint}. One can show that this is actually not the case, the reason is that our ensemble is just an infinite set of independent truncated Gaussian, indexed by the energy $E$. Now the moments of each of these truncated Gaussians are given by \eqref{TG1point} and \eqref{TGnpoint} and one can show using Carleman's condition that it is the unique distribution that reproduces these moments. We thank Arjun Kar for explaining to us this argument on the uniqueness of the distribution.}
\begin{equation}
\hat{\rho}(E)\ket{\rho}=\rho(E)\ket{\rho}\quad\text{with}\quad  \rho(E)\geq 0,\, \forall E.
\end{equation}
This has to be compared with the $\alpha$-states of the Gaussian model where $\rho(E)$ can take any value on the real axis. This shows that adding non-perturbative corrections can result in a drastic reduction of the baby universe Hilbert space.

\subsection{Non-perturbative completion}
\label{Non-perturbative completion}

The $\CGHS$ model has features that appear unphysical: a typical member of the ensemble has a continuous density of states. More worrisome is the fact that this density  can sometimes become negative. We have seen in the previous section that the negativity issue can be cured by non-perturbative corrections. But the problem remains that members in the ensemble have a continuous density of states and therefore do not admit a Hilbert space interpretation. In other words, the partition function for a typical member of the ensemble cannot be written as 
\be
Z(\b) =  \sum_{i=1}^L n_i\, e^{-\b E_i}~,
\ee
where $n_i$ are non-negative integers and the $E_i$'s are some discrete set of energies. In a physical ensemble we would expect all members to have a Hilbert space interpretation. A natural resolution of this problem is to view $\CGHS$ as resulting from the scaling limit of an ensemble of physical quantum systems.

This is precisely what happens in JT gravity, whose topological expansion is reproduced by the double-scaled limit of a matrix model. In that case, the member are physical quantum systems because their Hamiltonians are finite Hermitian matrices. One could wonder if such a matrix model description exists for $\CGHS$. 
 
 We have seen in Sec.~\ref{The Gaussian ensemble} that the two-point function of $\CGHS$ does not correspond the universal answer for a double-scaled matrix model. But this is assuming that our $Z(\beta)$ corresponds to $\Tr\,e^{-\b H}$, where $H$ is the matrix of the ensemble. One could imagine that the relation between our $Z(\beta)$ and the matrix is more involved. In Sec.~\ref{The Gaussian ensemble} we proposed a potential resolution for the absence of eigenvalue repulsion, relating it to the fact that we may not be working in the right thermodynamical ensemble. Now we make the following observation: if one starts with a true double-scaled matrix model, one has
\begin{equation}
 \langle \Tr\,e^{-\b_1 H} \Tr\,e^{-\b_2 H}\rangle_c\,\, \sim \,\,\frac{\sqrt{\b_1 \b_2}}{\b_1+\b_2},
\end{equation}
where the RHS corresponds to the universal answer for such ensembles. One could imagine fixing the charge associated with the gauge field at each boundary. In that case, the ground state energy would be shifted by $Q^2$ \cite{Iliesiu:2019lfc, Kapec:2019ecr}. This would produce the connected two-point function
\begin{equation}
 \langle \Tr\,e^{-\b_1 (H+Q_1^2)} \Tr\,e^{-\b_2 (H+Q_2^2)}\rangle_c\,\, \sim \,\,\frac{\sqrt{\b_1 \b_2}}{\b_1+\b_2}e^{-\b_1 Q_1^2-\b_2 Q_2^2}~.
\end{equation}
The absence of eigenvalue repulsion in our result would be a consequence of the fact that we are really at fixed chemical potential and therefore summing over all charge sectors. To fix the chemical potential, we should integrate over $Q_1$ and $Q_2$ which exactly cancels the $\sqrt{\b_1\b_2}$ factor and reproduce the $\CGHS$ answer \eqref{Zcylinder}. This suggests that $\CGHS$ should be dual to a double-scaled matrix model at fixed chemical potential. Indeed, suitable matrix models can be found \cite{WIP}.

\section{Flat wormhole solutions}\label{sec:WH}

In this section we consider $\CGHS$ gravity with matter fields, as described by the action
\be
I =I_{\mathrm{top}}+ I_{\CGHS} + I_\r{matter}+I_\partial~,
\ee
where the matter fields are not coupled to the dilaton. The boundary term $I_\partial$ is the boundary term considered in Sec.~\ref{The CGHS model}. Taking the matter to be a massless scalar,  we find a flat  Euclidean wormhole solution which shares many similarities with its JT gravity version described in \cite{Garcia-Garcia:2020ttf}. This wormhole dominates at low temperatures and we observe a phase transition from a connected wormhole phase to a disconnected phase with two black holes. 

 We also attempt to construct a Lorentzian eternal traversable wormhole similar to \cite{Maldacena:2018lmt}. This is done with a static coupling between the boundaries using $N$ massless scalar fields in the bulk. However, we find that $N$ must scales with the cutoff as $1/\e$ for the solution to exist. Hence, this Minkowski wormhole is not a solution of $\CGHS$. We encounter the same kind of difficulties that prevent simple constructions of traversable wormholes with Poincaré symmetry in higher-dimensional AdS \cite{Freivogel:2019lej}.


\subsection{Euclidean wormhole}\label{Euclidean wormhole}

We consider $\CGHS$ gravity coupled to a massless scalar field $\chi$. The matter action is
\be
I_\r{matter}={1\/2}\int d^2 x\sqrt{g} \,(\p\chi)^2~.
\ee
In Sec.~\ref{Euclidean path integral}, we consider the cylinder geometry. But this geometry is not a solution of the pure gravity action. Using a non-trivial  matter configuration, we would like to turn the cylinder geometry
\be
ds^2 = dt^2 + dx^2,\qq t\sim t+b~,
\ee
into a (classical) solution of the theory: a Euclidean wormhole. Note that in the path integral studied in Sec.~\ref{Euclidean path integral}, we integrate over all possible cylinder. This is not possible when we add a massless scalar field because of divergences at small $b$ due to its Casimir energy. The solution we construct allows to estimate the path integral in a saddle-point approximation, as will be illustrated by computing the quenched free energy of the wormhole.

The equations of motion for the metric and the gauge field are unchanged. The remaining equations of motion are
\bea
\n_\mu\n_\nu\Phi -g_\mn\Box\Phi  + \L g_\mn +  T_\mn^\chi\= 0~,\qq T_\mn^\chi = \p_\mu\chi \p_\nu\chi-{1\/2} g_\mn (\p\chi)^2~,\-
\Box\chi \=0~,
\eea
where we have set $\k= (8\pi G_2)^{-1} = 1$. We will consider a time-independent solution for the scalar field
\be
\chi  = \chi_1 x + \chi_0~,
\ee
where the constants will be fixed with appropriate boundary conditions. The solution for the dilaton takes the form
\be
\Phi = {1\/4}x^2 (2\L-\chi_1^2) + {1\/4}t^2 (2\L+\chi^2_1) + c_1 + c_2 t+c_3 x~.
\label{dilatonprofile}
\ee
The locus of the boundary of the spacetime is determined by the boundary condition on the dilaton
\be\label{WH:phibdy}
\Phi = {\phi_r\/\e}~.
\ee
Since we want the boundaries to be at $x=\r{const}$, we require $\Phi$ to be time-independent which gives the relation
\be
\chi_1  = i \sqrt{2\L}~,
\ee
and  leads to the solution
\be
\Phi = \L x^2~.
\ee
The integration constants are not important for the discussion so we set them to zero. From the boundary condition \eqref{WH:phibdy}, we find that the cutoff boundary is at
\be
x_c = \pm \sqrt{\phi_r\/\L\e}~.
\ee
We also have to impose the boundary condition on the gauge field.
\be
A|_\p = {d\tau\/\e} ~,
\ee
where $\tau\sim \tau+\beta$ is the boundary time. We introduce a coordinate system for the patch by defining
\be
t= {b\ov \beta}\tau +{\beta\ov b} i r,\qq x = {\b\ov b}r ~,
\ee 
so that the metric becomes
\be
ds^2 = {b^2\ov \b^2} d\tau^2 + 2 i d\tau dr ~.
\ee
We consider the solution $A=xdt+A_x (x)dx$ which is a solution to the equation $\epsilon^{\mn}\partial_\mu A_\nu=1$. Evaluating the solution at the cutoff boundary gives
\begin{equation}
A_\partial=\sqrt{\phi_r\/\L\e}\frac{\beta}{b}\,d\tau={d\tau\/\e} \,,
\end{equation}
and determines the size of the wormhole to be
\be\label{brelbetainWH}
b=\b \sqrt{\L\/\phi_r\e}~.
\ee
The last boundary condition is 
\be
(\Phi - i A^\mu \p_\mu\Phi)|_\p = \phi_h~,
\ee
where we have the freedom to choose the value of the constant $\phi_h$. This  condition gives
\begin{equation}
\Lambda (x_c^2-2i A_x(x_c) x_c)=\phi_h\,.
\end{equation}
so the boundary condition $\phi_h=0$ simply fixes $A_x=-i x/2$.\footnote{Any value of $\phi_h$ is actually possible by choosing a function satisfying $A_x \sim - i x/2 +i \phi_h/(2 x\Lambda)$ as $x\to+\infty$. }
For the cylinder, the value of $\phi_h$ is irrelevant since it doesn't enter in the effective action \eqref{onshellaction} as $P_0=0$ in the cylinder geometry.

At the cutoff boundary, the scalar field takes the value
\be\label{scalarbdycond}
\chi|_{x=\pm x_c} = \pm i \sqrt{2\phi_r\/\e}~.
\ee
These are the boundary conditions that we should impose on the scalar field to have the correct profile leading to the Euclidean wormhole solution. This should be interpreted as some kind of (imaginary) source for the scalar field, in the same way as the boundary sources used in AdS \cite{Garcia-Garcia:2020ttf, Marolf:2021kjc}. It is important that the sources are opposite on each side so that the scalar has a gradient in the bulk. As the analog of the AdS/CFT dictionary is not well-understood in flat spacetime, we cannot give a more precise interpretation of \eqref{scalarbdycond}.

The on-shell action of the wormhole takes the form
\be
I_\r{on-shell} = {\phi_r b^2 T} - 2 b\sqrt{\L\phi_r\/\e}~,
\label{Iwormhole}
\ee 
where $T$ is the temperature at each boundary.\footnote{We could have had different temperatures at each boundary by turning on the constant $c_3$ in the dilaton profile \eqref{dilatonprofile}. In that case, the formulae \eqref{brelbetainWH} and \eqref{Iwormhole} are the same with the replacement $T\to \frac{T_L+T_R}{2}$.} The first term is the contribution from the boundary action of $\CGHS$ computed in Sec.~\ref{Euclidean path integral} while the second term is the on-shell action of the scalar field. The result is obtained by plugging in the value \eqref{brelbetainWH} for $b$. If we consider instead this action for general $b$, one can check that \eqref{brelbetainWH}  is also the value that extremizes the action, which provides a nice consistency check of our boundary conditions.

To obtain a true solution of $\CGHS$ with matter, we should be able to take $\e\to0$ at the end. This is possible if we choose $\L = \e\,\L_0$  where $\L_0 > 0$ is some fixed constant. The cutoff can then be removed and we obtain
\be
b = \b\sqrt{ \L_0\/\phi_r}~,
\ee
with free energy
\be
F_\r{WH} = -{\L_0}~.
\ee
The constant free energy indicates a gapped system as expected for a wormhole \cite{Maldacena:2018lmt, Garcia-Garcia:2020ttf}.

\sss{Phase transition}

Let's now consider the solution from the boundary point of view. The Euclidean wormhole arises as a solution connecting two boundaries, called left (L) and right (R), where we impose the boundary condition
\be\label{chibdycond}
\chi|_{L}= i \sqrt{2\phi_r\/\e},\qq \chi|_{R} = -i\sqrt{2\phi_r\/\e}~.
\ee
Another solution with these boundary conditions consists of two black holes. In that case, the solution is just $\chi=\r{const}$ and the scalar field doesn't really do anything. The free energy of the two black holes is then
\be
F_\r{BH} = -2 T S_0~.
\ee
The physical free energy of the system with two boundaries is the minimum
\be
F = \r{min}(F_\r{WH} ,F_\r{BH})~.
\ee
We see that there is a phase transition at the temperature
\be
T_c = {\L_0\/2  S_0}~.
\ee
This is a phase transition between a disconnected phase with two black holes and a connected wormhole phase. Such a transition also occurs in similar wormhole solutions \cite{Garcia-Garcia:2020ttf, Marolf:2021kjc}.

\sss{Quantum effect of the matter}

\begin{figure}
  \centering
  \includegraphics[width=15cm]{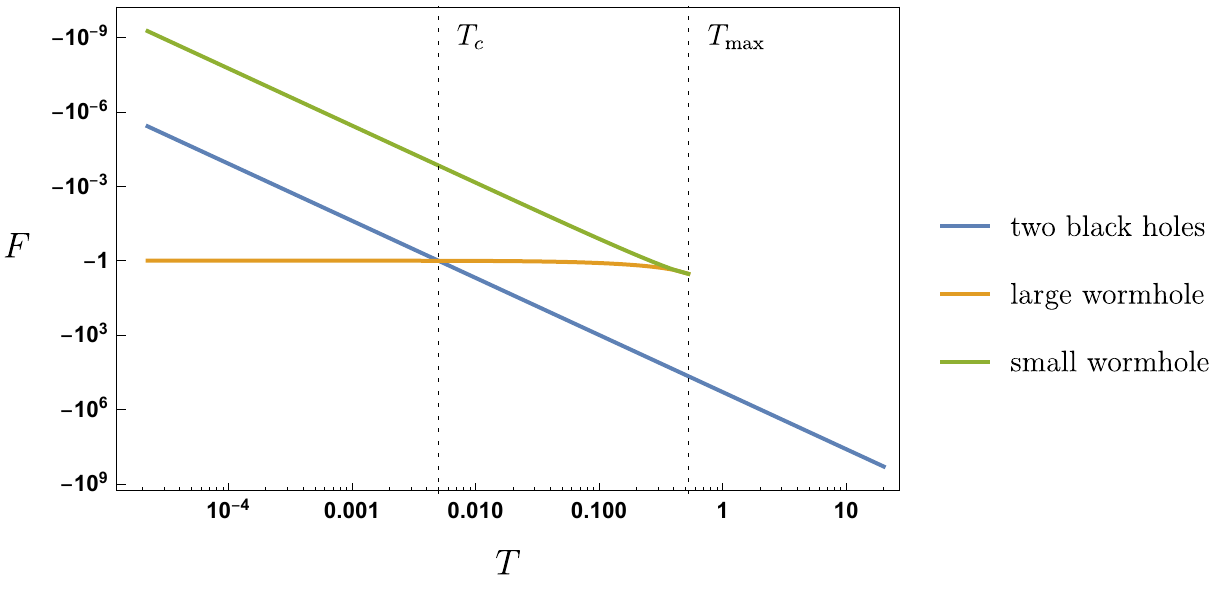}
  \caption{Log-log plot of the free energy. We observe a phase transition at $T=T_c$ from a disconnected phase with two black holes to the connected wormhole phase. The plot is done using $\L_0 =1$ and $S_0=10^2$.}\label{plotFtotal}
\end{figure}

\begin{figure}
  \centering
  \includegraphics[width=15cm]{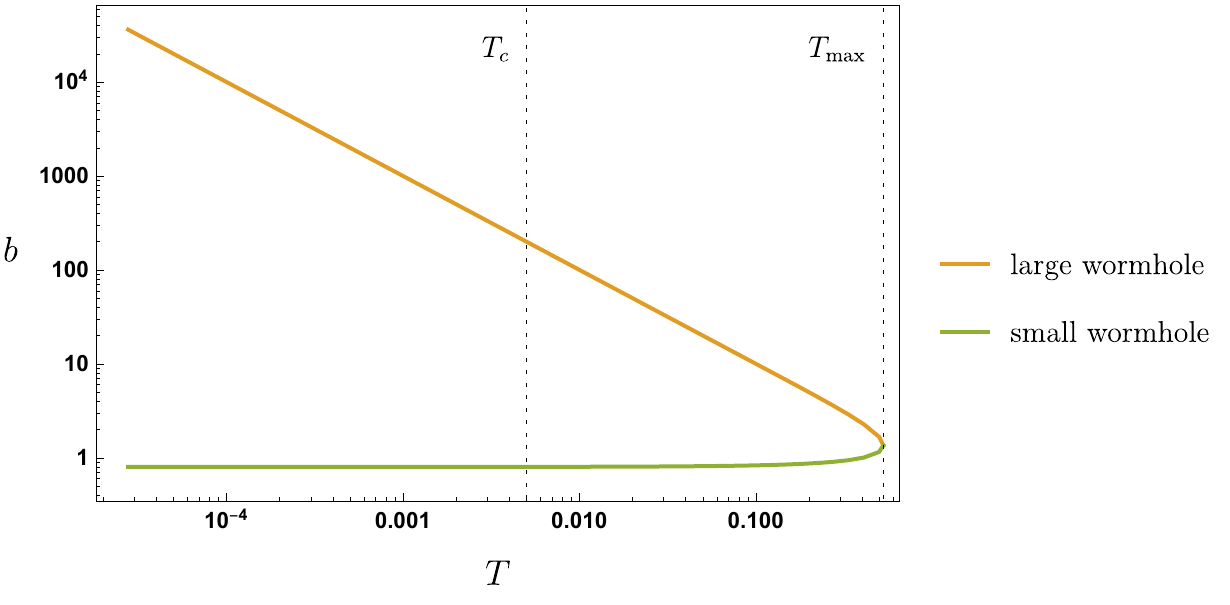}
  \caption{Log-log plot of the size $b$ of the wormhole as a function of the temperature. The plot is done using $\L_0 =1$ and $S_0=10^2$.}\label{plotSize}
\end{figure}

The above analysis was correct for large enough $b$ where the one-loop effects of the matter can be neglected. This effect simply adds to the on-shell action the term 
\be
{}-\log \Tr\, e^{-b L_0}~
\ee
which is the partition function of a 2d boson on the line at inverse temperature $b$. We will neglect the one-loop effect of the gravitational sector, giving a term proportional to $\log T$, which is only appreciable at very low temperatures. The resulting action turns out to be the same as the JT gravity version of this wormhole  so we won't repeat the analysis here.\footnote{In JT gravity, this answer is the result of two compensating effects: the Hamiltonian of the CFT on the strip has $L_0$ shifted by $-{c\/24}$ but this shift is cancelled by the Weyl anomaly as we go from the flat strip to AdS$_2$.} At the end, the temperature can be written as 
\be
T = {1\/b}\sqrt{\L_0\/\phi_r} - {\cE(b)\/2b}~,
\ee
where we have defined the positive function
\be
\cE(b) \equiv \r{Tr}(L_0 e^{-b L_0}) = {1\/24} + \p_b \log \eta(e^{-b})~.
\ee
This relation implies that there is a maximal value $T_\r{max}$ of the temperature above which the wormhole does to exist. At a temperature $T<T_\r{max}$, there are actually two wormhole solutions: a large wormhole described in the previous section but also a small wormhole at a much lower value of $b$. The complete phase diagram is plotted in Fig.~\ref{plotFtotal} and the relation between the size and the temperature are plotted in Fig.~\ref{plotSize}.

\sss{Quenched free energy of the wormhole}

\begin{figure}
  \centering
  \includegraphics[width=14cm]{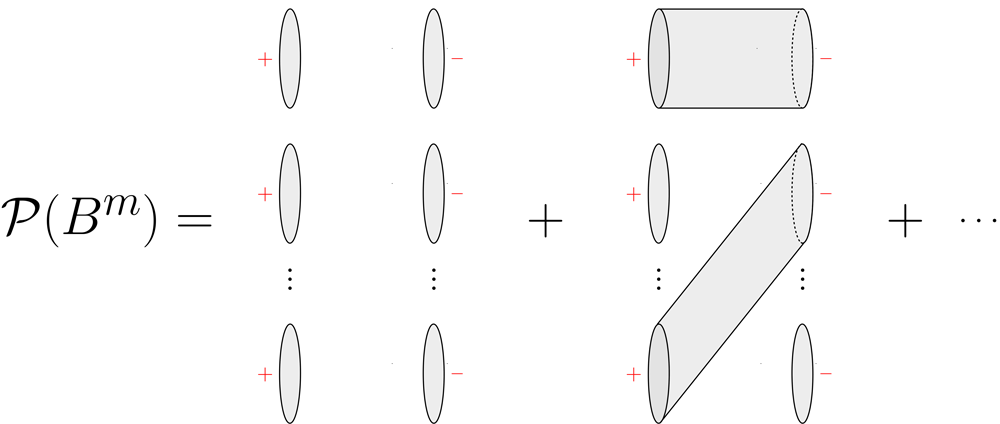}
  \caption{Computation of $\cP(B^m)$. Wormholes connect a $+$ boundary to a $-$ boundary. }\label{fig:Freplicas}
\end{figure}

In the above, we have computed the naive, \ie annealed, free energy $-T \log\ln Z\rn$ of the wormhole. This setup is simple enough that we can compute the quenched free energy $-T\ln \log Z\rn$ using replica wormholes \cite{Engelhardt:2020qpv}.  The idea is to write
\be
\ln \log Z(B)\rn = \lim_{m\to0} {1\/m}(\cP(B^m)-1)~,
\ee
where $B$ are the boundary conditions for our system. Here, we choose $B$ to consist of two circles: one labeled $+$ and the other $-$ corresponding to the two choices \eqref{chibdycond} of boundary conditions for the scalar field. The wormhole solution connects a $+$ circle to a $-$ circle. We can then evaluate $\cP(B^m)$ in saddle-point approximation using our wormhole solution, see Fig.~\ref{fig:Freplicas}. Under a natural analytic continuation in $m$, this yields
\be
\ln F\rn = F_\r{WH} + \g_\r{Euler} T~,\qq T<T_c~.
\ee
This is the same computation as in \cite{Garcia-Garcia:2020ttf} to which we refer to for more details. 

The correction from the replica wormholes corresponds to a shift in ground state entropy by $\g_\r{Euler} \approx 0.577$. Although it is interesting that such a correction can be computed, it turns out to be negligible here.

\subsection{Minkowski wormhole?}

In the previous section, we have constructed a Euclidean wormhole solution in $\CGHS$. This solution would appear in the path integral if we impose imaginary sources for the scalar field. It  naively doesn't  make sense in Lorentzian signature because  imaginary sources would allow arbitrarily negative energy and should be unphysical.

For these reasons, it would be desirable to have an eternal traversable wormhole in Lorentzian signature. Our goal is to construct a solution similar to that of \cite{Maldacena:2018lmt} in JT gravity. We attempt to do this using a double-trace coupling involving $N\gg 1$ massless scalar fields but we find no consistent wormhole solution. The difficulties we encounter are similar to the ones appearing when trying to construct  traversable wormholes in higher-dimensional AdS with Poincaré symmetry \cite{Freivogel:2019lej,VanRaamsdonk:2021qgv}, despite the fact that we are in two dimensions.

Let us consider the $\CGHS$ model with $N$ massless scalar fields
\be
I_\r{matter} = {1\/2}\sum_{i=1}^N\int d^2 x\sqrt{-g} \,(\p \chi_i)^2~.
\ee
We would like to find an eternal traversable wormhole which should be 2d Minkowski spacetime, described by the metric
\be
ds^2 = -dt^2 + dx^2~.
\ee
The two boundaries of the wormhole will be parallel and at
\be
x=\pm {L\ov 2}~,
\ee 
where  $L\to+\infty$ as we take the cutoff to zero. The boundary condition for the dilaton implies that we should have
\be\label{WHcondDilaton}
\Phi|_{x=\pm{L\/2}} = {\phi_r\/\e}~.
\ee
Without matter, the dilaton is monotonous in $x$ and cannot satisfies this condition. NEC-violating matter is required to obtain a consistent dilaton profile.

 For this reason, we introduce  a non-local coupling of Gao-Jafferis-Wall type \cite{Gao:2016bin}, between the two boundaries
\be
\d S = h \sum_{i=1}^N\int dt\,\chi_i\le(t,-\tfrac{L}{2}\ri) \chi_i\le(t,\tfrac{L}{2}\ri)~.
\ee
The resulting stress tensor is computed in \cite{Freivogel:2019lej} and takes the form 
\be
\ln T_\mn \rn_\r{coupling} \sim { h N\ov L}\bpm - 1 & 0 \\ 0 & -1 \epm~,
\ee
up to some numerical coefficient of order one. This computation is done in perturbation theory and requires that $h L\ll 1$. We should not forget that there is also the Casimir energy of the matter, which takes the form
\be
\ln T_\mn\rn_\r{Casimir} = {\pi N\ov 24 L^2}\bpm - 1 & 0 \\ 0 & -1 \epm~
\ee
for Dirichlet boundary conditions. We see that the Casimir energy is always dominating the contribution of the double-trace deformation. Increasing $h$ cannot work because even if we could go beyond perturbation theory in $hL$, the quantum inequalities \cite{Ford:1994bj} guarantees that we can at most get the effective value $h L \sim 1$, comparable to the Casimir contribution (see the discussion in \cite{Freivogel:2019lej}). The double-trace coupling is then irrelevant: we should only consider the Casimir energy and we can turn off the coupling.  This issue doesn't arise in AdS$_2$ because there is no Casimir energy there, the Casimir energy on the flat strip being cancelled by the Weyl anomaly.

We now solve the equation of motion for the dilaton
\be
\n_\mu\n_\nu\Phi -g_\mn\Box\Phi  +  g_\mn \L + \ln T_\mn\rn = 0~,
\ee
The general solution is
\be
\Phi = {1 \ov2}\le({\pi N\/24 L^2}-\L\ri)\,t^2+{1\/2}\le({\pi N\/24L^2}+\L\ri)\,x^2 ~,
\ee
where we have set some unimportant integration constants to zero. The condition \eqref{WHcondDilaton} implies that the dilaton is time-independent so that
\be
\L = {\pi N\/24 L^2}~.
\ee
We thus obtain the profile
\be
\Phi =\L x^2~. 
\ee
From \eqref{WHcondDilaton}, we obtain
\be
  N = {96\,\phi_r\/\pi\e}~.
\ee
This does not lead to a consistent solution as the limit $\e\to 0$ is singular, since the value of $N$ is fixed from the start and has to be finite. 

This conclusion was somewhat expected from the fact that the solution would be fully supported by Casimir energy, and traversable wormhole between two asymptotic boundaries should not exist if there is no coupling between the two boundaries. This follows from the no-transmission principle \cite{Engelhardt:2015gla}: a traversable wormhole between two asymptotically AdS regions has to rely on a coupling between the two dual CFTs (because the wormhole implies that they can  communicate through the bulk). This no-transmission  principle is one way to explain the difficulty in constructing  higher-dimensional AdS traversable wormhole with Poincaré symmetry \cite{Freivogel:2019lej}.\footnote{A  construction of such wormholes, using a rather different approach, was recently proposed \cite{VanRaamsdonk:2021qgv}. Relaxing the requirement of Poincaré symmetry, there is no issue in constructing eternal traversable wormholes between asymptotically AdS regions \cite{Bintanja:2021xfs}.  }  Here, we are discussing wormholes in flat spacetime but the no-transmission principle should still apply, as long as we believe in some notion of holography. 

\section{Discussion}

In this paper, we have shown that the $\CGHS$ model arises as a universal sector in the near horizon dynamics of non-extremal black holes. It would be interesting to understand in more details the physical significance of this near horizon limit and the interpretation of the $\CGHS$ sector from the far region. As there is no decoupling limit here, it seems that we should view the near horizon system as some kind of open system. 
It would also be interesting to understand the properties of this sector in rotating higher-dimensional black holes, such as the Kerr geometry, where the JT story is more intricate \cite{Castro:2019crn, Castro:2021csm}.

We have seen how the $\CGHS$ model, with its boundary action, correctly accounts for the change in entropy of a higher-dimensional black hole. This is captured by the disk geometry in two dimensions. On a more speculative note, one could wonder if the cylinder contribution has anything to say about the black hole. In the ensemble dual to $\CGHS$, the cylinder captures the eigenvalue statistics  which suggests that its higher-dimensional counterpart might know about statistical aspects of black hole microstates. See \cite{Cotler:2020ugk, Cotler:2020hgz,Belin:2020hea,Cotler:2020lxj} for recent progress  in the context of 3d gravity.



The $\CGHS$ model also provides a simple theory to study the gravitational path integral. The dual ensemble is Gaussian in the density of states. We expect that it should be corrected by non-perturbative contributions in some UV completion. In particular we have explained how such non-perturbative corrections can remove the members with negative density of states from the ensemble.

A lesson of our analysis concerns the puzzles about the quenched free energy in gravity. It was explained in \cite{Engelhardt:2020qpv} that the computation of the quenched free energy involves a replica trick with contributions from replica wormholes. It was proposed that these contributions might resolve an apparent unphysical feature of the annealed free energy, that it is non-monotonous. In this paper, we have computed the quenched free energy for $\CGHS$ directly, without having to deal with the issue of analytic continuation, and shown that it remains non-monotonous. As explained in Sec.~\ref{Quenched versus annealed free energy}, this suggests that the non-monotonicity has more to do with the fact that we are taking the continuum limit and that continuous systems can have a non-monotonous free energy. Moreover, our result for the quenched free energy of $\CGHS$ should be useful to understand the correct prescription for analytic continuation in the replica computation.

We expect that the $\CGHS$ model admits a matrix model description for reasons  explained in Sec.~\ref{Non-perturbative completion}. This would require  understanding  how to fix the $\r{U}(1)$ charge associated to the gauge field, which probably involve relaxing the additional boundary condition \eqref{CGHSadditionalcond}. It would be interesting to study this model and its dual matrix ensemble \cite{WIP}.

\begin{figure}
  \centering
  \includegraphics[width=7cm]{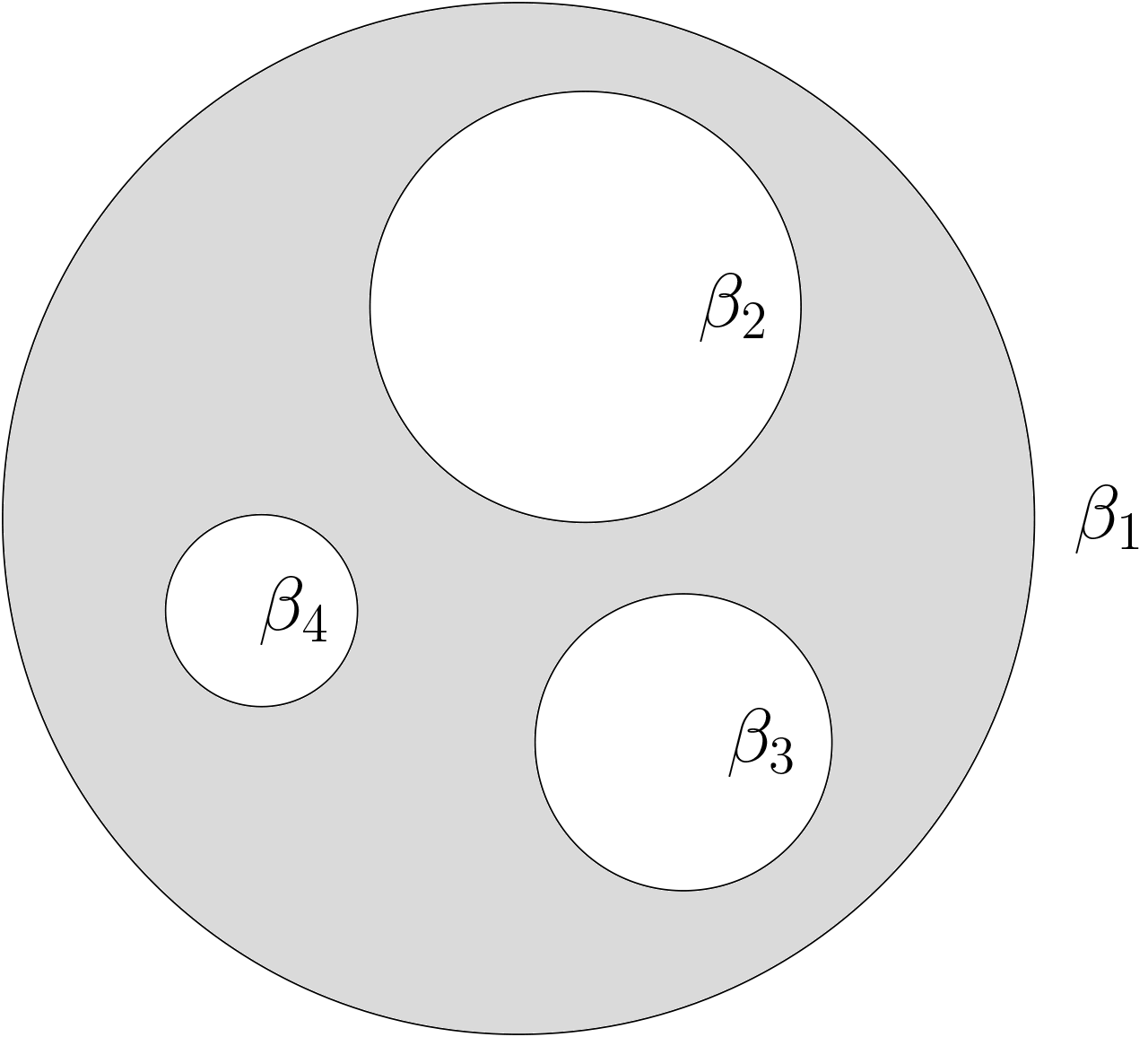}

  \caption{Example of a flat surface that would contribute to $\ln Z(\b_1)Z(\b_2)(Z(\b_3)Z(\b_4)\rn_c$ in the finite cutoff version of $\CGHS$. It would be interesting to define and study this theory.}\label{FigFC}
\end{figure}

Another interesting direction would be to study the theory at finite cutoff. In this theory, there would be flat surfaces connecting more than two boundaries, as depicted in Fig.~\ref{FigFC}. If such a theory makes sense, it would give an interesting model whose Euclidean path integral seems richer than $\CGHS$ while being more tractable than JT gravity at finite cutoff \cite{Iliesiu:2020zld}.

An aspect that is worth exploring is the existence of maximal chaos in the $\CGHS$ model. This should be a consequence of the near horizon symmetries in the bulk \cite{Shenker:2013pqa}. To show this, one could try to compute the appropriate out-of-time-ordered correlator on the boundary. This computation would require understanding better the dictionary between bulk and boundary operators in $\CGHS$. This could provide insights on flat holography.

The Euclidean wormhole that we constructed in Sec.~\ref{Euclidean wormhole}  deserves further study. It was argued in \cite{Garcia-Garcia:2020ttf} that the JT version of this Euclidean wormhole is dual to a two-site SYK model with complex couplings \cite{Garcia-Garcia:2021elz}. It would be interesting to see whether the Euclidean wormhole of $\CGHS$ has a similar interpretation, using the duality between $\CGHS$ and the complex SYK model proposed in \cite{Afshar:2019axx}.

We have given in this paper a covariant definition of $\CGHS$ using the boundary conditions \eqref{bdyCond}. We expect that similar boundary conditions can be used to give a covariant definition of the version of JT gravity studied in \cite{Godet:2020xpk} where new boundary conditions are imposed in addition to the usual ones. The covariant version of these boundary conditions would involve an additional gauge field, viewed as non-dynamical, which is used to fix the subleading component of the dilaton (see the App.~A of \cite{Godet:2020xpk} for the expression of the counterterm). Interestingly, these new boundary conditions for JT gravity go beyond the classification given in \cite{Goel:2020yxl}.

\section*{Acknowledgements}
We would like to thank Oliver Janssen, Arjun Kar and the rest of the string theory group at UBC for useful comments and discussion. VG acknowledges the postdoctoral program at ICTS for funding support through the Department of Atomic Energy, Government of India, under project no. RTI4001. CM acknowledges support from NSERC.

\appendix

\section{Example of non-monotonous free energy}
\label{From discrete to continuous}

In this Appendix, we present a simple  example of a quantum system whose free energy becomes non-monotonous in the continuous limit. We will construct a discrete model that reproduces the averaged partition function of $\CGHS$ in a continuous limit. In the discrete case, the free energy is always monotonous since the thermodynamical entropy is positive for all temperatures. In the continuum limit however, the free energy is not always monotonous, as  we know from Sec.~\ref{Quenched versus annealed free energy}, which implies that the thermodynamical entropy becomes negative below a certain critical temperature. 

We are looking for a discrete system whose density of states is linear in the continuum limit, see \eqref{pointfunctions}. A discrete system is characterized by its energy levels $E_k$ (with $E_{k+1}>E_k$) and the corresponding occupation numbers $n_k$. The partition function is given by 
\begin{equation}
Z_\r{discr}(\beta)=\sum_{k=1}^L n_k\, e^{-\beta\,  E_k}.
\label{DiscreteZ}
\end{equation}
Consider a system whose energy levels and occupation numbers are 
\begin{equation}
E_k=k\,\delta E,\quad n_k=e^{S_0}\,k.
\end{equation}
The energy levels are equally distributed and the occupation numbers are proportional to the energy.  One can compute its partition function exactly, the result is
\begin{equation}
Z_\r{discr}(\beta)=e^{S_0}e^{-\beta L \delta E}\frac{\left(e^{\beta (1+L)\delta E}+L-e^{\beta \delta E}(1+L)\right)}{\left(1-e^{\beta \delta E}\right)^2}.
\label{FreeDiscrete}
\end{equation}
For $T\to 0$, this is approximately $e^{S_0}e^{-\beta \delta E}$ while for $T\to\infty$, it becomes the constant $e^{S_0}{L(1+L)\/2}$. This implies the following behavior for the free energy
\begin{equation}
F(T) = \left\{
    \begin{array}{ll}
       \delta E-T S_0,\quad  & T\to 0,\\
       -T \log \left(\frac{1}{2}e^{S_0}L(1+L)\right),\quad & T\,\to \infty\,.
    \end{array}
\right.
\end{equation}
We see that when the temperature is small the free energy reaches a finite value that corresponds to the ground state energy. When the temperature is large, it is linear with a negative coefficient. One can see that the function \eqref{FreeDiscrete} is monotonic but this is guaranteed by the fact that the system is discrete.\footnote{The general proof goes as follows. Consider a system like \eqref{DiscreteZ}, the corresponding thermodynamical entropy $S=-\partial_T F$ is $S=\log Z+\beta\bar{E}$, where $\bar{E}$ is the mean value of the energy. Now using that $\log Z\geq \log n_1-\beta E_1 $ where $E_1$ is the smallest energy, and that $\bar{E}\geq E_1$, we conclude that $S\geq \log n_1$, where $n_1$ is the occupation number of the ground state. This guarantees that $S$ is always non-negative. The key point here was to be able to isolate the ground state which is not possible in a continuous system. }

\begin{figure}
  \centering
  \includegraphics[width=14cm]{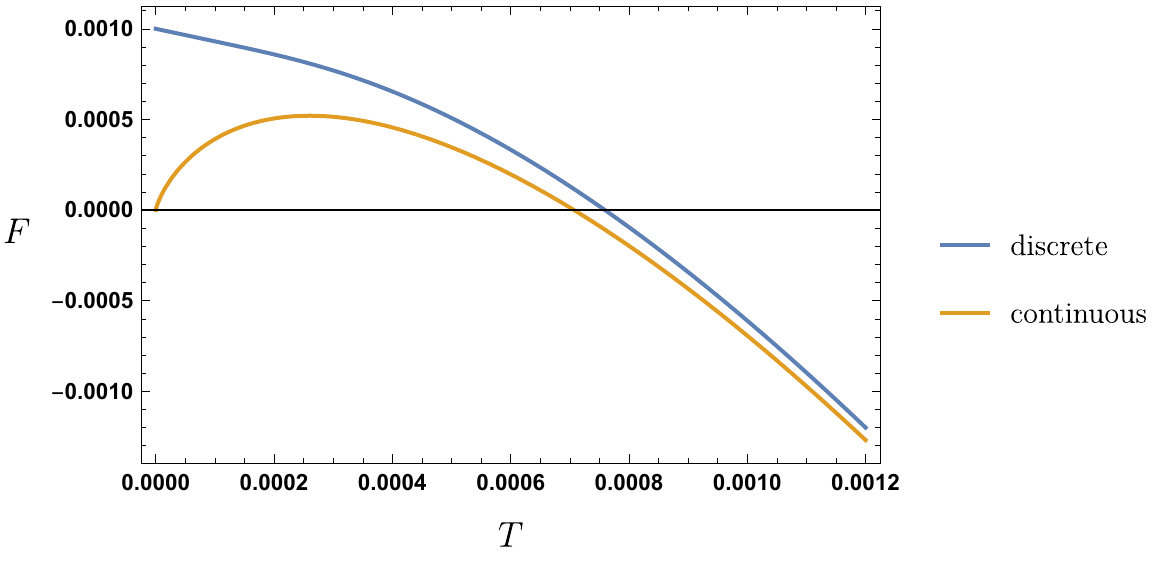}

  \caption{Free energies of a discrete system and its continuous limit. We see that a non-monotonous free energy can arise when taking the continuous limit of a discrete system. The plot was made with $e^{S_0}=2,\L=0.1$ and $L=100$.}\label{FigDiscrCont}
\end{figure}

We would like now to take a continuum limit $L\to \infty$. We also need to scale the energy step $\delta E$ with $L$ so that the spectrum becomes continuous. We take $\delta E = \Lambda/L$ so that the resulting continuous spectrum is the interval $E\in [0,\Lambda]$. The partition function becomes
\begin{equation}
Z_\r{discr}(\beta)=\frac{L^2}{\Lambda}\left[\frac{1}{L}\sum_{k=1}^L e^{S_0}E_k\, e^{-\beta\,  E_k}\right].
\end{equation}
According to standard results on Riemannian integration, the sum in the bracket converges to the integral $\frac{1}{\Lambda}\int_0^\Lambda e^{S_0}E \,e^{-\beta E}$. Therefore we define the continuous partition function to be
\begin{equation}
Z_{\mathrm{cont}}(\beta)\equiv 2\pi \lim_{L\to \infty}\left(\frac{\Lambda}{L}\right)^2 Z_\r{discr}(\beta),
\end{equation}
where the limit is taken at fixed $\Lambda$. The rescaling between the discrete and continuous partition functions can be seen as the addition of a constant term to the action of the continuous theory. The factor $2\pi$ is here just to have the right normalization. The final  result is
\begin{equation}
Z_{\mathrm{cont}}(\beta)=\frac{2\pi \,e^{S_0}}{\beta^2} \left(1-(1+\beta\Lambda)e^{-\beta \Lambda}\right).
\end{equation}
One can check that the corresponding continuous free energy is non-monotonous for any value of the cutoff $\Lambda$, see Fig.~\ref{FigDiscrCont}. Taking the infinite cutoff limit, we recover the averaged partition function of the $\CGHS$ model \eqref{correlZ} which lead to a non-monotonous free energy, with negative thermodynamic entropy in the region $T< e^{S_0/2}/\sqrt{2\pi}$. Therefore we see that the continuum limit has the effect of making the free energy non-monotonous.

\bibliography{bibliography}

\end{document}